\documentclass[final,3p,times,twocolumn]{elsarticle}

\usepackage{amssymb}
\usepackage{amsmath}
\usepackage{threeparttable}  

\usepackage{natbib}
\journal{Physics of the Dark Universe}

\begin{document}

\begin{frontmatter}



\title{What is the amount of baryonic dark matter in galaxies?}


\author {V\' aclav Vavry\v cuk}

\affiliation{organization={Charles University, Faculty of Science},
            addressline={Albertov 6}, 
            city={Prague},
            postcode={12800}, 
            country={Czech Republic, },
            email = {email: vavrycuk@natur.cuni.cz}
            }

\begin{abstract}
In this paper, we re-evaluate the estimates of dust mass in galaxies and demonstrate that current dust models are incomplete and based on a priori assumptions. These models suffer from a circularity problem and account for only a small portion of dust, specifically submicron-sized grains. They overlook larger dust particles and other macroscopic bodies, despite observational evidence supporting their existence. This evidence includes the observed (sub)millimeter excess in dust emission spectra and the power-law size distribution with a differential size index $\gamma \approx 3.5-4.0$, which has been measured for large particles and compact bodies across diverse environments. Examples of these large particles include large dust grains and meteoroids detected by satellites, near-Earth objects colliding with Earth, fragments in the Main Asteroid Belt and the Kuiper Belt, interstellar 'Oumuamua-like objects, and exoplanets. As a result, dust-type baryonic dark matter may be more abundant throughout the galaxy by one order of magnitude or even more than previously assumed, with a significant portion of its mass concentrated in large compact bodies. Additionally, black holes may contribute significantly to the total mass of baryonic dark matter. Consequently, current galaxy models do not provide reliable estimates of baryonic mass in galaxies. Clearly, a substantially larger amount of baryonic dark matter in galaxies would have major implications for theories of galaxy dynamics and evolution.

\end{abstract}


\end{frontmatter}
 
\section{Introduction}

The existence of dark matter was first postulated by Zwicky \citep{Zwicky1933, Zwicky1937}, who analyzed the unexpectedly high orbital velocities of galaxies in clusters. These velocities were inconsistent with the gravitational potential generated solely by the observed luminous matter, suggesting a significant amount of unseen or missing mass. This ‘missing mass’ problem was later corroborated by observations of the flat rotation curves of disk galaxies. According to Newton's gravitation law, the tangential velocity of stars should decrease with distance from the galaxy centre. However, observations show that in most cases, the velocity remains nearly constant. The discovery of flat rotation curves was first made by Rubin and Ford \citep{Rubin_Ford1970} for the Andromeda galaxy and was later confirmed for other spiral galaxies \citep{Rubin1980,Rubin1985, Albada1985, Kent1987, Begeman1989, Persic1996, Sanders1996, Sofue_Rubin2001, deBlok2002}.

Originally, Zwicky \citep{Zwicky1937} suggested that dark matter consisted of dust, microscopic and macroscopic solid bodies, and gas. However, the baryonic origin of dark matter was later questioned and largely dismissed. Analyses of galaxy rotation curves indicate that the mass of dark matter might be much higher than the mass of dust and gas in galaxies \citep{Albada1985, Dubinski_Carlberg1991, Navarro1996, Navarro1997, Persic1996}. Consequently, dark matter was considered as primarily non-baryonic \citep{White_Rees1978, Davis1985, White1987, Maddox1990a, Moore1999, Bergstrom2000, DelPopolo2013, Bertone_Hooper2018}. However, the existence of non-baryonic dark matter is often criticized as physically controversial \citep{Kroupa2012, Kroupa2015, Del-Popolo_Le-Delliou2017}. Consequently, alternative theories have been proposed to explain flat rotation curves without non-baryonic dark matter \citep{Milgrom1983a, Milgrom2012, Bekenstein2004, Mannheim2012, Mannheim2019, Vavrycuk_Frontiers_Astron_Space_Sci_2023}.

The problem of baryonic dark matter in galaxies is, however, more complex than previously thought. Analyses of galaxy spectra reveal an unexpected excess in far-infrared (FIR), submillimetre and millimetre dust emission (known as the 'submm excess'). This excess is primarily observed in the spectral energy distributions (SEDs) of many low-metallicity galaxies \citep{Galliano2005, Ade2011_Planck_XVII_Submm_excess, Galametz2011, Dale2012, Remy-Ruyer2013, Hermelo2016, Dale2017, Turner2019}, indicating a clear discrepancy between dust models and observations. Furthermore, satellite and other measurements of dust and micrometeorites in the Solar System \citep{Grun2001, Strub2015}, along with studies of asteroid distribution in the Main Asteroid Belt in the Solar System \citep{Ivezic2001, Jedicke2002, Bottke2005a, Gladman2009}, show that the dust mass distribution in the solar neighbourhood differs significantly from that considered by standard dust models \citep{Mathis1977} and their modifications \citep{Draine_Lee1984, Kim1994, Weingartner_Draine2001, Draine2003b, Zubko2004}. 

In this paper, we address these discrepancies by revisiting current dust models and their mass estimates for dust and other cold bodies in galaxies. We show that these models are incomplete and based on \textit{a priori} assumptions, thus tracing only a small portion of  dust-type baryonic dark matter. Consequently, they do not provide reliable estimates of the total dust mass in galaxies. We propose an alternative model, called the Cold-Body (CB) model, which assumes that the dust-type baryonic dark matter is an order of magnitude or more higher than previously estimated. This model is consistent with current observations of the submm excess in dust emission spectra, as well as with observations of large interplanetary and interstellar particles and bodies. Clearly, a considerably greater amount of baryonic dark matter in galaxies would have substantial implications for theories of galaxy dynamics and evolution.

\section{Observations of interstellar dust}

Dust is an important component of the interstellar and intergalactic medium, playing a key role in processes ranging from gas chemistry and thermodynamics to star formation \citep{Draine2003, Draine2011}. Dust grains form in the rapidly cooling gas of stellar outflows \citep{Draine_Salpeter1979a, Draine_Salpeter1979b, Dominik_Tielens1997, Dwek1998}, emerging from asymptotic giant branch (AGB) stars, and from nova or supernova (SN) explosions. The typical size of dust grains is less than 1 $\mu$m, although larger grains are also observed. Dust grains can growth by coagulation or accretion in dense clouds, but they may also be destroyed through thermal sputtering, SN shock waves, and collisional shattering \citep{Jones1996}. 

The details of the evolution process of grains are still uncertain, but it is clear that the growth of small-sized particles into larger aggregates must be robust and rapid \citep{Hirashita_Kuo2011, McKinnon2016} to produce not only millimetre-sized grains, as indicated by radio observations of protoplanetary disks \citep{Andrews_Williams2005, Rodmann2006, van-der-Marel2013, Kataoka2014}, but also larger compact particles of sub-meter scale or even planetesimals, asteroids, and planets \citep{Weidenschilling2011} as detected by the Sloan Digital Sky Survey \citep{Ivezic2001}, the Wide-field Infrared Survey Explorer \citep{Mainzer2011}, and other surveys \citep{Tedesco2002, Brown2018}.

Submicron dust grains are assumed to form needle-shaped, irregular fractal, or fluffy aggregates, which are often highly porous \citep{Wright1987, Henning1995, Dominik_Tielens1997, Kataoka2014}. As grain size increases, porosity decreases, and dust particles become compact and rocky \citep{Dominik_Tielens1997, Blum_Wurm2008, Chiang_Youdin2010}. Dust consists of a mixture of amorphous carbon, silicates, graphite, and polycyclic aromatic hydrocarbons (PAH) \citep{Draine2003}. Dust grains are electrically conductive and cause wavelength-dependent extinction characterized by the power law $A_V \sim \lambda^{-\beta}$, with a dust emissivity index $\beta$ ranging between 1.4 and 2. Additionally, strong extinction features at 9.7 $\mu$m and 18 $\mu$m caused by silicates, a hydrocarbon feature at 3.4 $\mu$m, and PAH absorption at 6.2 $\mu$m and 7.7 $\mu$m are observed \citep{Draine2011}. Conductivity and the elongated shape of small dust grains causes that light emitted by dust is polarized according to the magnetic field in galaxies \citep{Ade2015_Planck_XIX_Polarized_emission}. Moreover, wavelength-dependent extinction produces reddening of light measured by the reddening coefficient $R_V = A_V/(A_B-A_V)$, where $A_B$ ad $A_V$ are the $B$-band and $V$-band extinctions, respectively. The mean value  is $R_V \approx 3.1$ for the Milky Way but can vary, being larger along lines of sight of toward dense clouds (e.g., $R_V \approx 5.5$ toward the Orion nebula), see Cardelli et al. \citep{Cardelli1989} or Fitzpatrick et al. \citep{Fitzpatrick1999}. 

Dimming and reddening due to dust significantly affect galaxy light at the UV to NIR wavelengths, with more than 30\% of starlight in galaxies absorbed by dust and re-emitted at infrared wavelengths. This percentage, however, depends on galaxy type and age. Elliptical galaxies show rather low effective extinction $A_V \approx 0.04-0.08$, while irregular and spiral galaxies exhibit higher extinction $A_V \approx 0.3-0.95$ \citep{Calzetti2001}. Detailed studies indicate that arms in the galaxy disks are more opaque, with mean values  $A_V \approx 1.4-5.5$ and peak values $A_V \approx 10-12$, while regions between arms are more transparent with $A_V \approx 0.3-1.1$ \citep{Beckman1996, Holwerda2005a, Holwerda2005b}. Interestingly, high-redshift galaxies also contain significant amount of dust \citep{Pettini1994, Ledoux2002}; dusty, evolved galaxies at $z > 7$ have been reported by Watson et al. \citep{Watson2015} and Laporte et al. \citep{Laporte2017}. 

\section{Models of dust in galaxies and their verification}

Detailed information about stars, gas and dust in galaxies is contained in the galaxy spectral energy distributions (SEDs). The amount of dust, its spatial distribution, and temperature can be determined using theoretical models that solve the radiative transfer equation \citep{Witt1992, Silva1998, Misselt2001, Tuffs2004, Honig2006, Popescu2000, Popescu2011, Walcher2011} or through semi-analytic approaches that use calibrated spectral libraries to interpret infrared emission \citep{Cunha2008}. To reproduce the SEDs at FIR wavelengths, a mixture of warm and cold grains with temperatures of 30-60 K and 15-25 K, respectively, is often considered \citep{Calzetti2000, Cunha2008, Galametz2012}.

\begin{figure*}
\centering
\includegraphics[angle=0,width=14.0 cm,trim=00 50 00 20]{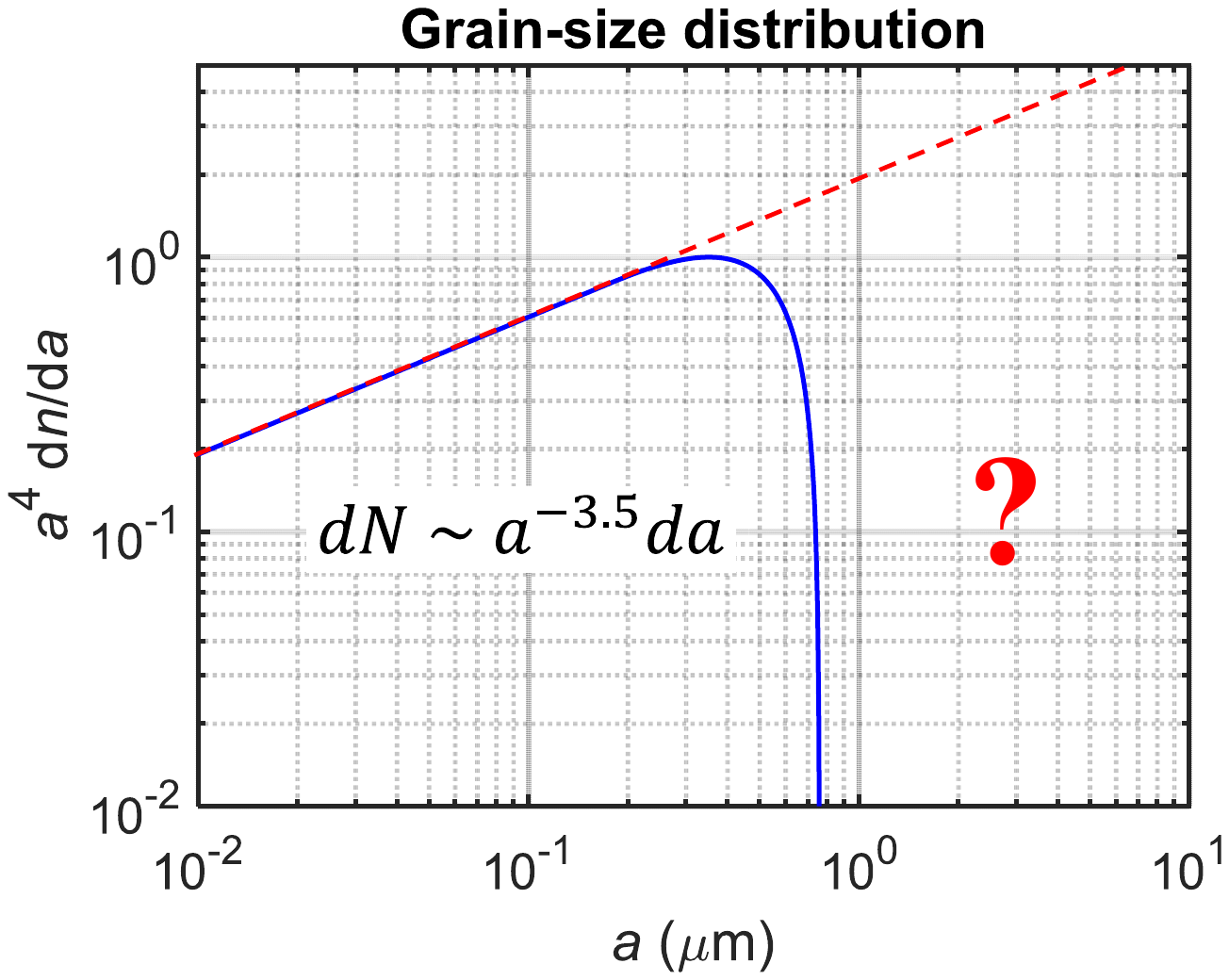}
\caption{
 The grain-size distribution of the MRN model. The dashed red line represents the power law with a size index $\gamma = 3.5$; the solid blue line represents the power law with a sharp cutoff applied in the MRN model \citep{Mathis1977}. The distribution function is normalized to its maximum value. The question mark emphasizes that the model fails to account for grains larger than 1 $\mu$m.
}
\label{fig:1}
\end{figure*}

\subsection{The MRN dust model}

The majority of codes modelling SEDs use the so-called MRN dust model \citep{Mathis1977} or its modifications \citep{Draine_Lee1984, Kim1994, Weingartner_Draine2001, Draine2003a, Draine2003b, Zubko2004}. The MRN model is defined by a power-law grain-size distribution $N(a)$ 
\begin{equation}\label{eq1}
dN \sim a^{-\gamma}da \,, \,\, a_{\mathrm{min}} < a < a_{\mathrm{max}} \,,
\end{equation}
where $dN$ is the number density of spherical grains with radii $a$ within the interval $(a, a+da)$, and $N$ is the total number of grains with radii less than $a$. The slope $\gamma$ is 3.5, known as the differential size index. The lower and upper size limits in Equation (1) are usually considered as $a_{\mathrm{min}} \approx 0.005$ $\mu$m and $a_{\mathrm{max}} \approx 0.25$ $\mu$m (see Figure~\ref{fig:1}). The size index $\gamma$ also controls the grain-mass distribution $N(m)$
\begin{equation}\label{eq2}
dN \sim m^{-s}dm\,, \, s=\frac{\gamma+2}{3} \,,
\end{equation}
where $dN$ is he number density of grains within the mass interval $(m, m+dm)$, and $s$ is known as the differential mass index. The mass-size distribution $M(a)$ is defined as
\begin{equation}\label{eq3}
dM \sim a^{3-\gamma} da \,, \, M(a) \sim a^{4-\gamma} \,,
\end{equation}
where $dM$ is the total mass of grains with radii $a$ within the interval $(a, a+da)$, and $M(a)$ is the total mass of grains with radii smaller than $a$. Hence, for $\gamma < 4$, most of the mass is concentrated in largest particles.

If a mixture of graphite and silicate particles is considered, the size index $\gamma$ and the size limits $a_{\mathrm{min}}$ and $a_{\mathrm{max}}$ may differ slightly for each component. The Mie theory and dielectric functions for graphite and silicate particles are used to compute extinction cross sections and, subsequently, the wavelength-dependent extinction. Parameters of the dust model are optimized by fitting predicted extinction functions to observations \citep{Mathis1977, Draine_Lee1984, Weingartner_Draine2001}. More complex models incorporate ultra-small PAH particles with irregular shapes and effective radii $a < 50$ \AA $\,$ to fit IR emission features at 6.2 $\mu$m and 7.7 $\mu$m \citep{Li_Draine2001b, Draine_Li2007}. This dust model, with minor modifications, is widely used in studies modelling dust absorption/emission in galaxies \citep{Silva1998, Popescu2000, Tuffs2004, Zubko2004, Cunha2008, Popescu2011, Inoue2020}.

\subsection{Estimates of dust mass and dust-to-gas mass ratio in galaxies}

The total dust mass $M_D$ of radiating dust clouds is often determined using the FIR and submillimetre continuum emission from dust grains. At these wavelengths, the emission is optically thin, meaning radiative transfer corrections are unnecessary, and the dust mass can be estimated from the following formula \citep{Hildebrand1977, Hildebrand1983, Draine2011}
\begin{equation}\label{eq4}
M_D = \frac{F_{\lambda} R^2}{\kappa_D B_\lambda (T)} \,, \,\, 
\kappa_D = \frac{3}{4} \frac{Q_\lambda}{a \rho} \,,
\end{equation}
where $F_\lambda$ is the flux density at wavelength $\lambda$, $R$ is the distance to the cloud, $B_\lambda (T)$ is the Planck function describing the radiation of a blackbody at temperature $T$, $\kappa_D$ is the mass opacity (or mass absorption coefficient), $\rho$ is the specific density of dust, $a$ is the radius of the dust grains, and $Q_\lambda$ is the absorption coefficient (or emissivity at long wavelengths) of dust at wavelength $\lambda$. In the geometric limit, where $a \gg \lambda/2 \pi$, the absorption coefficient is $Q_\lambda \approx 1$. In the Rayleigh limit, where $a \ll \lambda/2 \pi$, coefficient $Q_\lambda$ decreases as $\sim 1/\lambda$ for amorphous grains and as $\sim 1/\lambda^{2}$ for crystalline grains \citep{Ossenkopf1992}. 

A key factor in determining dust mass is the mass opacity $\kappa_D$ defined in Equation (4). For example, Li and Drain \citep[in their table 6,]{Li_Draine2001b} report a mass opacity of 0.043 kg$^{-1}$ m$^{2}$ at $\lambda = 850 \, \mu$m, although other values are also used: 0.057 kg$^{-1}$ m$^{2}$ \citep{Baes2020} or 0.077 kg$^{-1}$ m$^{2}$ \citep{Dunne2000, Cunha2008}. Since $\kappa_D$ depends on the specific density $\rho$ of grains, which is affected by unknown porosity, and on the absorption coefficient $Q_\lambda$, whose decay with $\lambda$ at Rayleigh limits is not precisely known, the true value of $\kappa_D$ is uncertain and may easily differ by an order of magnitude from the values used in modelling.

An alternative approach is to determine the dust-to-gas mass ratio and to estimate dust mass by measuring the total amount of gas in the cloud or galaxy. Bohlin et al. \citep{Bohlin1978} found that the ratio of the total hydrogen column density to the colour excess $E(B-V)$ is approximately constant, with a value of $5.8 \times 10^{21} \, \mathrm{cm}^{-2}$. By fitting the extinction curve with various mixtures of silicate and carbonaceous dust grains, it is possible to estimate the total volumes $V_S$ and $V_G$ of silicate and carbonaceous grain populations per H atom. For example, Weingartner and Draine \citep[in their case B,]{Weingartner_Draine2001} report $V_S = 3.9 \times 10^{-27} \, \mathrm{cm}^{3} \, \mathrm{H}^{-1}$ for the silicate grains and $V_G = 2.3 \times 10^{-27} \, \mathrm{cm}^{3} \, \mathrm{H}^{-1}$ for the carbonaceous grains, assuming $R_V = 3.1$. This yields a dust-to-gas mass ratio of 0.01 for the Milky Way. Similar or lower values are reported by other authors and for various galaxies \citep{Draine2007, Munoz-Mateos2009, Draine2014, Watson2015, Aniano2020}. Some studies also indicate a positive correlation between the dust-to-gas mass ratio and the galaxy metallicity \citep{Lisenfeld_Ferrara1998, Galametz2011, Sandstrom2013}. 

\subsection{Limitations of the MRN dust model}

The MRN dust model and its modifications (e.g., the WD01 model \citep{Weingartner_Draine2001} in Table 1) apply a sharp cutoff to the maximum grain size at $a_{\mathrm{max}} \approx 0.1-0.5 \, \mu$m (see Figure~\ref{fig:1}). Clearly, strictly limiting dust grains to sizes below 1 $\mu$m is questionable and contradicts observations of microscopic grains and macroscopic solid bodies with a broad range of sizes, detected particularly in circumstellar disks at FIR and (sub)millimetre wavelengths \citep{Wilner2005, Rodmann2006, Wyatt2008, Birnstiel2016, Portegies-Zwart2018, Manser2019}. The constraint on the maximum limit $a_{\mathrm{max}} < 1 \, \mu$m in current dust models is largely formal, based on mathematical rather than physical considerations, as the conditions for dust grain formation and growth in the interstellar medium (ISM) remain uncertain \citep{Dwek1998, Hirashita_Kuo2011, McKinnon2016}. 

The strict size limit reflects the fact that absorption and emission of larger dust grains is weak or even negligible compared to small grains. Hence, determining the distribution of larger grains with $a_{\mathrm{max}} > 1 \, \mu$m from their absorption/emission properties is a difficult and ill-conditioned task. In addition, interstellar dust is commonly studied as part of gas clouds, which are rich in small grains. While small grains are well coupled with the gas, large particles begin to decouple from gas flow, as their surface-to-mass ratio decreases with particle growth \citep{Birnstiel2016}. Consequently, dust particles not bounded to gas are mostly ignored in current dust models.

\begin{figure*}
\centering
\includegraphics[angle=0, width=15.0 cm, trim=60 0 10 10]{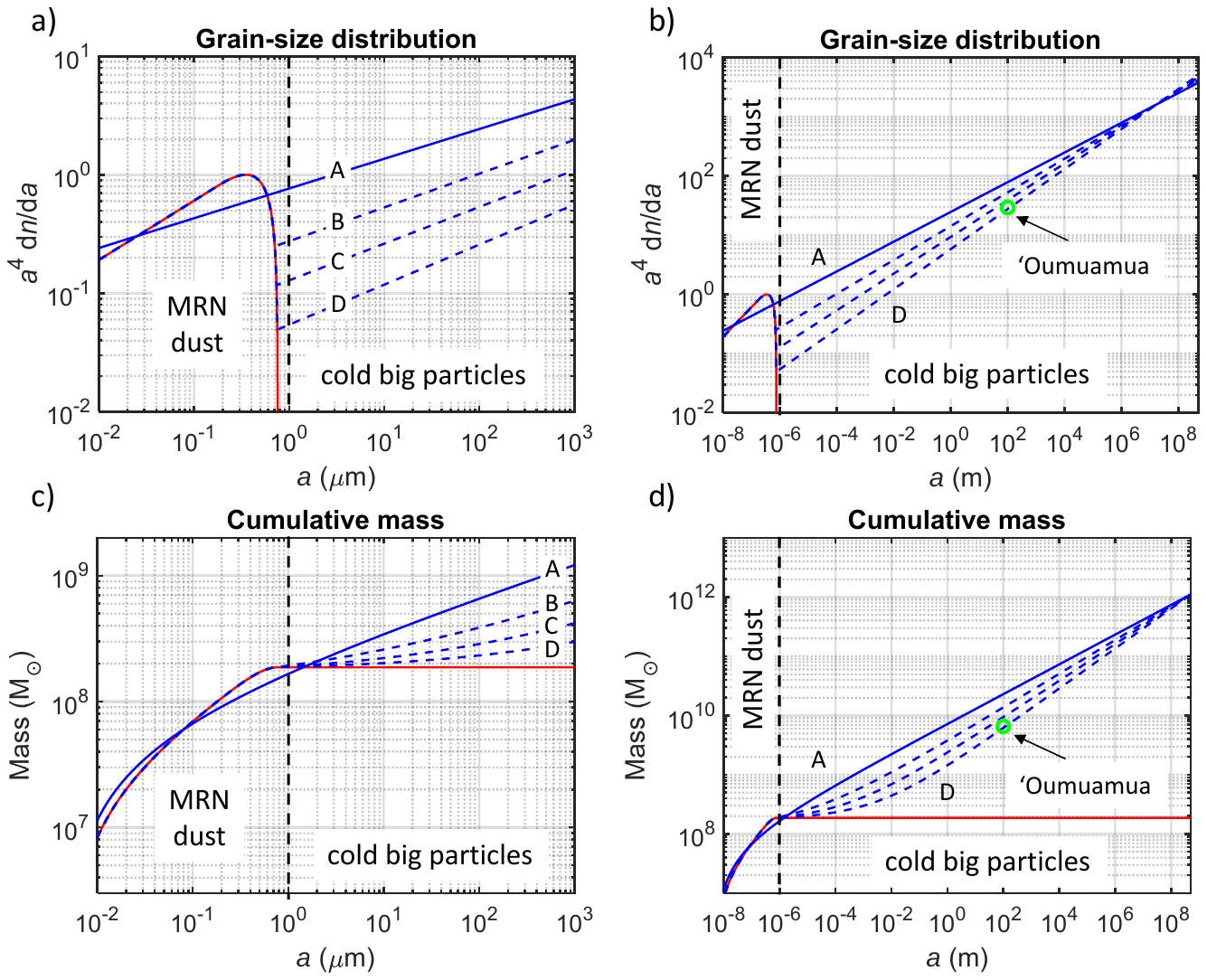}
\caption{
The grain-size distribution and cumulative mass of the CB model are shown for grain size $a$ less than 1 mm (a,c) and for the full size range of $a$ (b,d). The solid blue line represents the CB-A model, while the dashed blue lines represent the CB-B, CB-C, and CB-D models. The solid red line corresponds to the MRN model. The green circle indicates the number density and mass of 'Oumuamua-like interstellar objects, as reported by Do et al. \citep{Do2018}. Plot (d) demonstrates that considering dust grains and particles with size $a > 10^{-6}$ m can lead to dust mass estimates (blue lines) that are even several orders higher than current estimates (red line). If we assume a size index $\gamma$ higher than that for the CB-A model, ranging between 3.75 and 4.0, the cumulative dust mass continues to increase with $a$, but at a gentler slope. Consequently, the total dust mass will be lower than the upper limit of $10^{12} M_\odot$ used in plot (d). For model parameters, see Table~\ref{Table:1}. 
}
\label{fig:2}
\end{figure*}

\begin{table}
%
\centering
%
%
\caption{Parameters of extinction law for dust models}  

\label{Table:1}      
%
\begin{tabular}{|l|l|l|l|l|l|}  
%
%
%
\hline                 
%
%
 Model & WD01 & CB-A & CB-B & CB-C &  CB-D\\
%
%
\hline                        
%
%
Size index $\gamma$         &  3.50 & 3.75 & 3.72 & 3.69 & 3.66 \\
$R_V$                       &  3.16 & 2.97 & 3.09 & 3.05 & 3.03 \\    
$\kappa_{1.1}/\kappa_{300}$ &  2760 &  312 &  661 &  985 & 1420 \\    
$\kappa_{1.1}/\kappa_{850}$ & 19700 & 1450 & 3110 & 4790 & 7440 \\  
$\beta_{300}$               &  2.15 & 1.54 & 1.57 & 1.62 & 1.71 \\  
$\beta_{850}$               &  1.69 & 1.26 & 1.26 & 1.27 & 1.31 \\  
%
%
\hline                                  
\end{tabular}
%
%
\begin{tablenotes}

\item {WD01 - modified MRN model proposed by Weingartner and Draine \citep{Weingartner_Draine2001}, $R_V$ is the reddening ratio, $\kappa_{1.1}/\kappa_{300}$ and $\kappa_{1.1}/\kappa_{850}$ are the ratios of the emission optical depths, and $\beta_{300}$ and $\beta_{850}$ are the slopes of the extinction law at 300 and 850 $\mu$m. Note that $\gamma = 3.5$ for WD01 is an overall approximate value, because this model considers slightly different size indices for individual dust components. }

\end{tablenotes}
%
\end{table}

\section{Considering large dust particles and compact bodies}

The limitations of current dust models can be addressed by introducing a more general 'Cold-Body model' (CB model), which incorporates cold large particles and macroscopic bodies into the MRN model. The CB model can be constructed in several alternative ways. First, we can simply assume that the grain-size distribution described by Equation (1) applies across the entire range of possible particle sizes
\begin{equation}\label{eq5}
dN \sim a^{-\gamma}da \,, \,\, 5 \times 10^{-9} < a < 5 \times 10^{8} \, \mathrm{m} \,.
\end{equation}
The uniform size index $\gamma$ in Equation (5) can then be determined by satisfying two conditions: (1) the dust mass for grains with sizes $0.005 \, \mu\mathrm{m} < a < 1 \, \mu\mathrm{m}$ matches the dust mass predicted by the MRN model, and (2) the total dust mass equals the mass of dark matter estimated for our Galaxy (Figure~\ref{fig:2}, blue solid lines). 

Alternatively, we can assume a combination of two distinct grain-size distributions: (1) the MRN distribution for $a < 1 \, \mu$m, and (2) a different grain-size distribution for $a > 1 \, \mu$m: 
\begin{equation}\label{eq6}
dN_1 \sim a^{-3.5}da \,, \,\, 5 \times 10^{-9} < a < 2.5 \times 10^{-7} \, \mathrm{m} \,,
\end{equation}
\begin{equation}\label{eq7}
dN_2 \sim a^{-\gamma}da \,, \,\, 1 \times 10^{-6} < a < 5 \times 10^{8} \, \mathrm{m} \,,
\end{equation}
\begin{equation}\label{eq8}
dN = dN_1 + dN_2 \,.
\end{equation}
In this case, the size index $\gamma$ in Equation (7) is calculated for the population of large grains to ensure the dust mass aligns with the estimated mass of dark matter in the Galaxy (Figure~\ref{fig:2}, blue dashed lines). The three alternative distributions shown in Figure~\ref{fig:2} simulate models characterized by a low number density of dust grains in the range from 1 $\mu$m to 100 $\mu$m. 

\begin{figure*}
\centering

\includegraphics[angle=0,width=14.5 cm,trim=0 40 40 20]{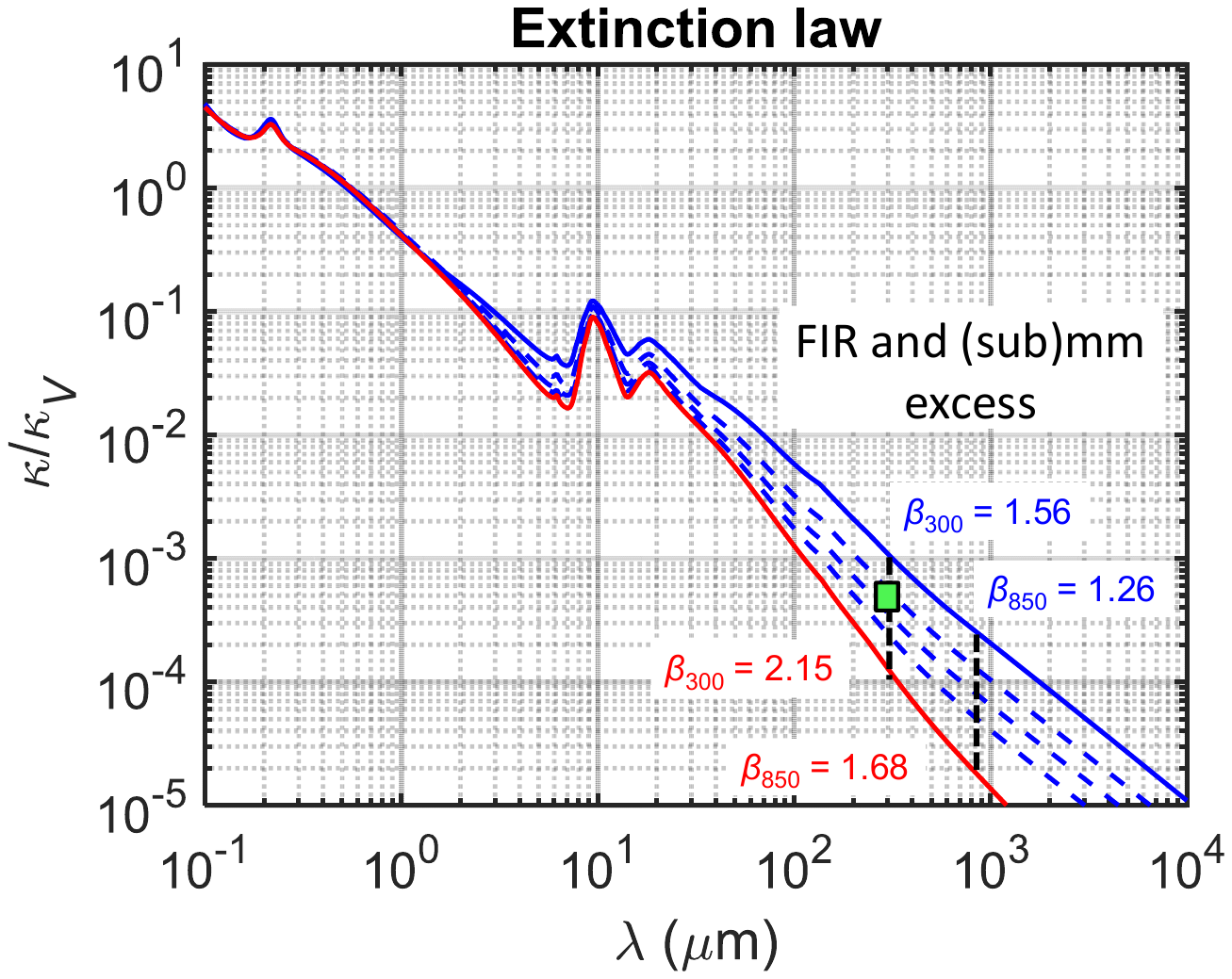}
\caption{
The extinction law for the standard dust model (red line) and the CB models (blue lines). The red line represents the WD01 dust model (Weingartner and Draine, ApJ, 2001); the solid blue line represents the CB-A model, and the dashed blue lines represent the CB-B, CB-C, and CB-D models. For model parameters, see Table~\ref{Table:1}. The dashed black lines mark extinction at wavelengths of 300 and 850 $\mu$m. The green rectangle marks the extinction derived from the observed ratio of emission optical depths $\kappa_{1.1}/\kappa_{300}$ for the M31 galaxy as reported by Whitworth et al. \citep{Whitworth2019}. The figure demonstrates: (1) the FIR and (sub)mm excess, evidenced by the visibly higher extinction of the CB models compared to the WD01 model for $\lambda > 20 \, \mu$m, and (2) the lower values of slope $\beta$ of the CB models compared to the WD01 model for $\lambda > 20 \, \mu$m. 
}
\label{fig:3}
\end{figure*}

Table~\ref{Table:1} shows that the size index $\gamma$ is higher for the modified dust models compared to the WD01 model but does not exceed a value of 4. Thus, the majority of the mass remains concentrated in the largest particles. Although the total dust mass is substantial in the modified dust models, the impact on the extinction law is weak or moderate (see Figure~\ref{fig:3}). The larger the particle size, the lower the opacity of these particles. For dust models with a reduced number density of particles between 1 $\mu$m and 100 $\mu$m, the reddening ratio $R_V$ is almost unaffected (see Figure~\ref{fig:3} and Table~\ref{Table:1}). Visible differences appear only at (sub)millimetre or larger wavelengths, where the CB models predict about an order of magnitude higher extinction and a slope $\beta \approx 1.5-1.6$ at a wavelength of 300 $\mu$m, which is lower than $\beta \approx 2.1-2.2$ predicted by the WD01 model. The CB models also predict a different behaviour for the temperature of dust grains. While small dust grains exhibit higher temperatures and act as grey bodies due to their limited emissivity at long wavelengths, large particles are colder and should emit nearly as blackbodies.

\section{Observations of the FIR and (sub)millimetre excess}

The excess in FIR, submillimetre and millimetre dust emission (referred to 'submm excess') predicted by the numerical modelling described above has been observationally confirmed by many researchers. This excess is primarily observed in the SEDs of many low-metallicity galaxies \citep{Galliano2005, Ade2011_Planck_XVII_Submm_excess, Galametz2011, Dale2012, Remy-Ruyer2013, Hermelo2016, Dale2017, Turner2019}. The studies show that the emissivity spectra  flatten, particularly, in the submm and mm regions. However, a discrepancy between models and observations is also evident in the FIR region. For example, Whitworth et al. \citep{Whitworth2019} report that the observed ratio of the emission optical depths $\kappa_{1.1}/\kappa_{300}$ for the M31 galaxy is 3-5 times lower than predicted by various theoretical dust models (see Figure~\ref{fig:3}, green square) and argue that current dust models need revision. 

Additionally, many researchers report values of the spectral emissivity index $\beta \approx 1.2-1.7$ for cold dust, similar to the slope in Figure~\ref{fig:3} and Table~\ref{Table:1} predicted by the CB model, being significantly lower than $\beta \approx 2.0$ predicted by standard dust models \citep{Ade2011_Planck_XVII_Submm_excess, Aghanim2016_Planck_XLVIII_Dust_emission, Gordon2010, Tabatabaei2014}. The SED flatness at long wavelengths requires the presence of very cold grains with temperatures $T \approx 5-9$ K, likely formed by dust aggregation, and indicates a substantial increase in total dust mass, with more than 50\% residing in these very cold grains \citep{Lisenfeld2002, Galliano2003, Galliano2005, Galametz2009, Galametz2011, Paradis2009}. Thus, the submm excess strongly supports the proposed CB model.

Although the hypothesis of very cold grains with a large total mass is frequently discussed as a possible explanation for the submm excess, it is often rejected by arguments that such a dust model predicts an excessively large dust mass and an unrealistic dust-to-gas mass ratio \citep{Galliano2011, Ade2011_Planck_XVII_Submm_excess, Hermelo2016}. Other frequently proposed mechanisms for the submm excess include: dust grains composed of amorphous carbon rather than graphite \citep{Meixner2010}, spinning dust \citep{Hermelo2016}, changes in the dust emissivity spectral index due to anomalous intrinsic properties of dust  \citep{Galliano2011}, or the cosmic microwave background (CMB) anomaly \citep{Bot2010}. However, these theories also face criticism and remain controversial on certain points \citep{Bot2010, Hermelo2016, Mason2020}, and no explanation has yet gained widespread acceptance.

\section{Observations of large interplanetary and interstellar particles and bodies}

Independent information on the size distribution of dust grains and particles in the ISM can be gathered from observations within the Solar System and other circumstellar disks, where a mixed population of particles, ranging from dust grains to large bodies such as planetesimals or planets, is observed \citep{Grun2019, Koschny2019}. In such environments,  dust grains must grow by at least 15 orders of magnitude in size, indicating that the formation of large bodies is likely a robust mechanism \citep{Birnstiel2016}. Although the number density of dust grains in circumstellar disks is several order of magnitude higher than the average density in the interstellar medium, the statistical properties of dust grains might be similar (Table~\ref{Table:2}). Moreover, these systems are not isolated, as they interact with the ISM. During planet formation, a considerable portion of the mass from planetary systems is ejected into interstellar space \citep{Dones2004, Fitzsimmons2018}.

\subsection{Satellite and other measurements of dust and micrometeorites}

In situ measurements of interstellar dust fluxes within the Solar System were conducted by the Ulysses dust detector from 1992 to 2007 \citep{Grun2001, Strub2015}. The Ulysses data reveal that dust mass distribution at the solar neighbourhood differs significantly from that of the MRN model \citep{Grun2001, Frisch1999, Kruger_Grun2009, Kruger2015}. Specifically, large dust grains with mass exceeding 10$^{-13}$ kg (diameter $a > 2 \, \mu \mathrm{m}$) have been detected, while smaller dust grains were observed in much lower quantities than predicted by the MRN model. The Ulysses dust measurements were later confirmed by the Galileo and Cassini spacecrafts \citep{Altobelli2005, Altobelli2007}. Other satellites also measured the flux and size distribution of particles with $a > 10 \, \mu$m. For example, Merouane et al. \citep{Merouane2016} analysed data from comet 67P/Churyumov-Gerasimenko using the COSIMA instrument on board Rosetta, finding a cumulative size distribution of dust with index of $1.9 \pm 0.3$ for dust particles between 30 $\mu$m and 150 $\mu$m. Similarly, Love and Brownlee \citep[their fig. 1] {Love_Brownlee1993} used the Long Duration Exposure Facility satellite to study the meteoroid flux, observing a size index from of $1.5$ to $3.0$ for dust particles between 50 $\mu$m and 300 $\mu$m. Also, radar-detected meteoroids of the Solar System dust cloud detected large grains (from 30 $\mu$m to 2 mm) with a differential mass index of approximately $2.0-2.1$, corresponding to $\gamma = 4.0-4.3$ \citep{Galligan_Baggaley2004, Pokorny_Brown2016}. Finally, similar results were obtained by Jewitt et al. \citep{Jewitt2014}, who used the Hubble Space Telescope data for studying dust size distribution in the tail of comet 133P/ELST-PIZARRO, finding a differential size index in the range from 3.25 to 3.5 for particle sizes between $0.2$ mm and $10$ mm.

\subsection{Near-Earth objects colliding with the Earth}

Asteroids with  $a  \lesssim 100$ m can be studied by measuring the flux of near-Earth objects colliding with Earth. Brown et al. \citep{Brown2002} report that the flux of objects in the $1-10$ m size range follows the same power-law distribution as bodies larger than 50 m. Based on various datasets, such as satellite data, infrasonic/acoustically measured bolide flux \citep{ReVelle2001}, Lunar cratering flux \citep{Werner2002}, and Near-Earth Asteroid Tracking and Spacewatch surveys \citep{Rabinowitz2000}, the cumulative size index was determined to be 2.7, corresponding to a differential size index $\gamma$ of 3.7. As shown by Bland and Artemieva \citep{Bland_Artemieva2006}, this value holds even when data are extended to cm-sized asteroids using the Canadian camera network atmosphere data \citep{Halliday1989}.

\subsection{Asteroids, planetesimals and exoplanets}

The size distribution of asteroids has been also studied through observations of fragments in the Main Asteroid Belt in the Solar System. This research is based on surveys such as the Sloan Digital Sky Survey (SDSS) and the Sub-Kilometer Asteroid Diameter Survey (SKADS) \citep{Ivezic2001, Jedicke2002, Bottke2005a, Gladman2009} and covers a size range from hundreds of metres to hundreds of kilometres. Bottke et al. \citep{Bottke2005a} report a size index ranging from 2.3 to 4.0 with an average value of 3.5.

Additionally, Artemieva et al. \citep{Ananyeva2020} analysed the mass distribution of 210 exoplanets detected and confirmed by the Kepler space telescope, as well as over 300 transiting giant exoplanets observed by ground-based surveys and the CoRoT space telescope. They studied planets with masses between 0.02 and 13 Jupiter masses, and found a mass distribution described by a power law, $dN/dm \sim m^{-s}$. The mass index $s$ is between 1.9 and 2.1, translating to a size index $\gamma$ between 3.7 and 4.3 (see Equation 2).

\begin{figure*}
\centering

\includegraphics[angle=0,width=14.5 cm, trim= 0 50 0 20]{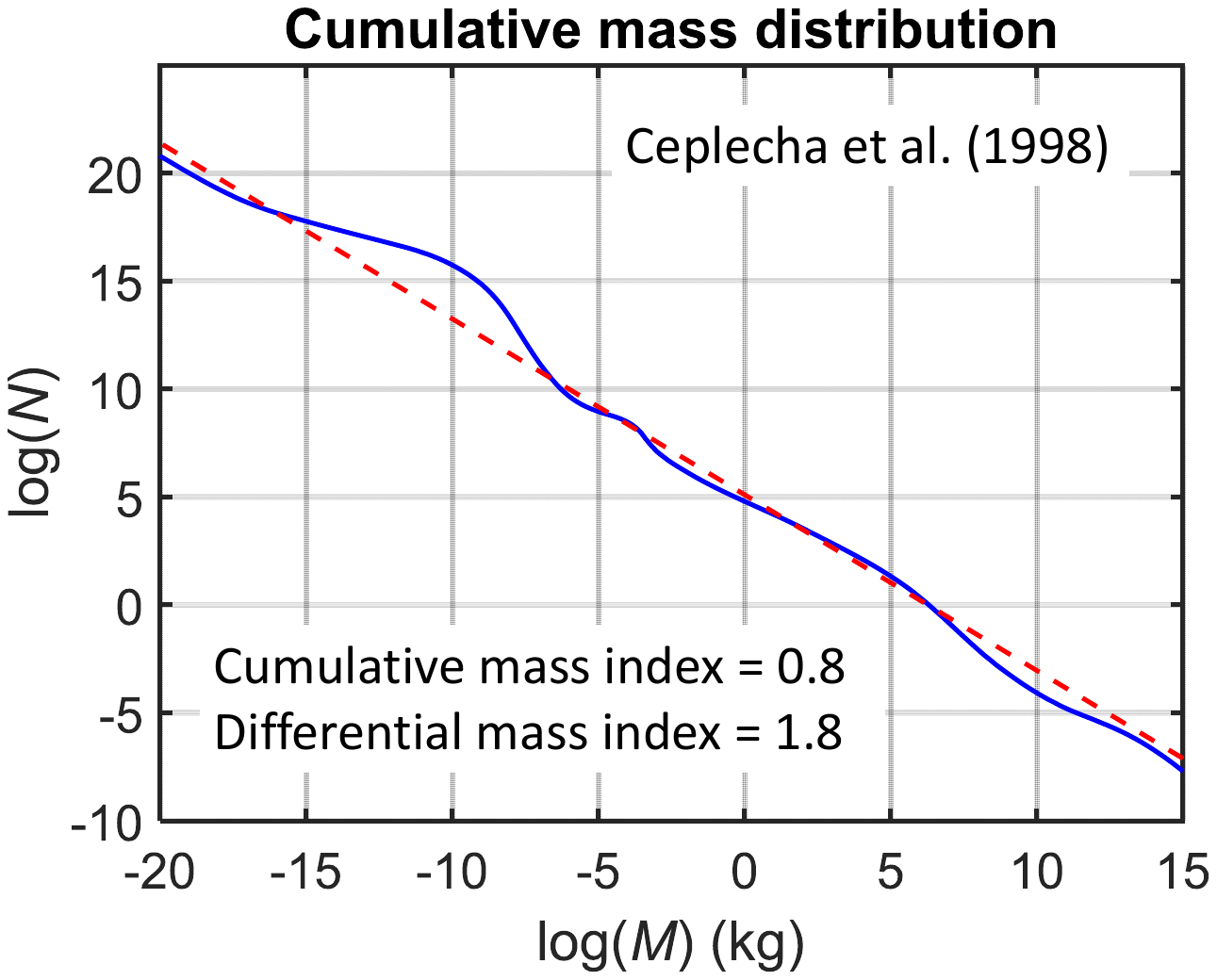}
\caption{
The cumulative mass distribution of interplanetary bodies. The blue line represents the observed mass distribution as reported by Ceplecha et al. \citep{Ceplecha1998}, while the red dashed line represents the power-law distribution with a differential mass index $s = 1.8$, corresponding to a size index $\gamma = 3.4$ (see Table~\ref{Table:2}).
}
\label{fig:4}
\end{figure*}

\subsection{Large interstellar objects}

The discovery of the interstellar object 'Oumuamua, observed by the Pan-STARRS1 telescope on  October 19, 2017, provided evidence of large, cold interstellar objects not gravitationally bound to any star \citep{Meech2017, Bolin2018, Trilling2018}. Characterized by an extremely elongated shape with a mean radius of 102 m and by a hyperbolic orbit as it passed through our Solar System, it radically changed our understanding of interstellar object density in the Galaxy. Estimates place the number density of the 'Oumuamua-like objects at $n_{\mathrm{obs}} \approx 2 \times 10^{15} \, \mathrm{pc}^{-3}$ \citep{Do2018, Portegies-Zwart2018}, suggesting that such objects could indeed form outside of stellar systems. The high observed density contrasts with previous assumptions that large objects are unlikely to form in interstellar space. The estimated high number density of 'Oumuamua-like objects points to their formation in the interstellar space, as the ejection of material from planetary systems into interstellar space during planet formation \citep{Dones2004, Fitzsimmons2018} is insufficient to populate the Galaxy by such objects \citep{Do2018}.

If large particles can indeed evolve in interstellar space, they might follow a similar power law in particle size distribution as found in circumstellar disks. Consequently, cold large particles and bodies can be ubiquitous in interstellar space. Assuming a Galaxy radius of 26 kpc and a width of 1.2 kpc, a rough estimate of the number density of 'Oumuamua-like asteroids predicted by the CB model is in the range $n_{\mathrm{theor}} \approx 1.9-5.7 \times 10^{15} \, \mathrm{pc}^{-3}$, which aligns well with the density $n_{\mathrm{obs}} \approx 2 \times 10^{15} \, \mathrm{pc}^{-3}$ obtained from the 'Oumuamua observation. Thus, the discovery of 'Oumuamua provides key observational support for the existence of ambient baryonic dark matter in the galaxy, with a size distribution consistent with the CB model (see Figure~\ref{fig:2}b,d, green circle).


\begin{table*}
%
\centering
%
%
\caption{Differential mass- and size-distribution index for the MRN model and observational dust data}  

\label{Table:2}      
%

\resizebox{\textwidth}{!}{%
\begin{tabular}{|c|c|c|c|c|c|c|c|}  
    
%
%
\hline\hline                 
%
%
\multicolumn{2}{|c|}{Mass range} & \multicolumn{2}{|c|}{Size range} & \multicolumn{2}{|c|}{Index} & & \\
$m_\mathrm{min}$ & $m_\mathrm{max}$ & $a_\mathrm{min}$ & $a_\mathrm{max}$ & Mass & Size & Method/Data/Model & Authors \\
(kg) & (kg) & (m)  & (m) & $s$ & $\gamma$ & & \\
%
%
\hline                        
\multicolumn{8}{|c|}{Interstellar dust models} \\
\hline                        
%
%
$ 1.3 \times 10^{-21} $ & $ 1.6 \times 10^{-16} $ & $ 5 \times 10^{-9} $ & $ 2.5 \times 10^{-7} $ & 1.8 & 3.5 & MRN model & Mathis et al. \citep{Mathis1977} \\
%
%
$ 1.3 \times 10^{-21} $ & $ 1.3 \times 10^{30} $ & $ 5 \times 10^{-9} $ & $ 5 \times 10^{8} $ & 1.92 & 3.75 & CB-A & this paper \\
%
%
$ 1.3 \times 10^{-21} $ & $ 1.6 \times 10^{-16} $ & $ 5 \times 10^{-9} $ & $ 2.5 \times 10^{-7} $ & 1.8 & 3.5 & CB-B & this paper \\
$ 3.6 \times 10^{-15} $ & $ 1.3 \times 10^{30} $ & $ 7 \times 10^{-7} $ & $ 5 \times 10^{8} $ & 1.91 & 3.72 & &  \\ 
%
%
$ 1.3 \times 10^{-21} $ & $ 1.6 \times 10^{-16} $ & $ 5 \times 10^{-9} $ & $ 2.5 \times 10^{-7} $ & 1.8 & 3.5 & CB-C & this paper \\
$ 3.6 \times 10^{-15} $ & $ 1.3 \times 10^{30} $ & $ 7 \times 10^{-7} $ & $ 5 \times 10^{8} $ & 1.90 & 3.69 & &  \\ 
%
%
$ 1.3 \times 10^{-21} $ & $ 1.6 \times 10^{-16} $ & $ 5 \times 10^{-9} $ & $ 2.5 \times 10^{-7} $ & 1.8 & 3.5 & CB-D & this paper \\
$ 3.6 \times 10^{-15} $ & $ 1.3 \times 10^{30} $ & $ 7 \times 10^{-7} $ & $ 5 \times 10^{8} $ & 1.89 & 3.66 & &  \\ 
\hline                        
\multicolumn{8}{|c|}{Measurements of interplanetary dust and bodies} \\
\hline                        
$ 3.5 \times 10^{-11} $ & $ 4.4 \times 10^{-9} $ & $ 1.5 \times 10^{-5} $ & $ 7.5 \times 10^{-5} $ & 1.5-1.7 & 2.6-3.2 & Satellite comet & Merouane et al. \citep{Merouane2016} \\
 & &  &  &  & & data &  \\
$ 1.3 \times 10^{-9} $ & $ 2.8 \times 10^{-7} $ & $ 5 \times 10^{-5} $ & $ 3 \times 10^{-4} $ & 1.5-2.0 & 2.5-4.0 & Satellite data & Love and Brownlee \citep{Love_Brownlee1993} \\
$ 3 \times 10^{-10} $ & $ 1 \times 10^{-4} $ & $ 3 \times 10^{-5} $ & $ 2 \times 10^{-3} $ & 2.0-2.1 & 4.0-4.3 & Radar-detected & Galligan and Baggaley \citep{Galligan_Baggaley2004}, 
 \\
 & &  &  &  & & meteor data & Pokorny and Brown \citep{Pokorny_Brown2016}
 \\
$ 8 \times 10^{-8} $ & $ 1 \times 10^{-2} $ & $ 2 \times 10^{-4} $ & $ 1 \times 10^{-2} $ & 1.75-1.83 & 3.25-3.5 & HST main-belt & Jewitt et al. \citep{Jewitt2014}
 \\
 & &  &  &  & & comet data & 
 \\
$ 1 \times 10^{-20} $ & $ 1 \times 10^{15} $ & $ 1 \times 10^{-8} $ & $ 5 \times 10^{3} $ & 1.8 & 3.4 & Flux data on & Ceplecha et al. \citep{Ceplecha1998}
 \\
 & &  &  &  & &  Earth-crossing & 
 \\
 & &  &  &  & & asteroids & 
 \\
$ 1.0 \times 10^{-1} $ & $ 1.0 \times 10^{10} $ & $ 0.5 $ & $ 100 $ & 1.9 & 3.7 & Asteroids colliding  & Brown et al. \citep{Brown2002}
 \\
 & &  &  &  & &  with the Earth & Halliday et al. \citep{Halliday1989}
 \\
$ 7.4 \times 10^{11} $ & $ 1.6 \times 10^{20} $ & $ 500 $ & $ 3 \times 10^{5} $ & 1.8 & 3.5 & Main Asteroid  & Ivezic et al. \citep{Ivezic2001} 
 \\
 & &  &  & (1.4-2.0) &(2.3-4.0) &  Belt & Jedicke et al. \citep{Jedicke2002}
 \\
 & &  &  &  & &   & Bottke et al. \citep{Bottke2005a}
 \\
 & &  &  &  & &   & Gladman et al. \citep{Gladman2009}
 \\
$ 3.5 \times 10^{16} $ & $ 1.1 \times 10^{20} $ & $ 1.5 \times 10^{4} $ & $ 2.2 \times 10^{5} $ & 2 & 4 & Kuiper Asteroid  & Schlichting et al. \citep{Schlichting2013},  
 \\
 & &  &  & & &  Belt & Fuentes et al. \citep{Fuentes2010}
 \\
$ 3.8 \times 10^{25} $ & $ 2.5 \times 10^{28} $ & $ 1.5 \times 10^{7} $ & $ 1.3 \times 10^{8} $ & 1.9-2.1 & 3.7-4.3 & Exoplanets & Ananyeva et al. \citep{Ananyeva2020} 
 \\
%
%
\hline                                  

\end{tabular}
}
%
%
\begin{tablenotes}
\item {HST - Hubble Space Telescope. The mass-size limits are recalculated under the assumption of spherical particle shapes and a specific mass density of 2500 kg m$^{-3}$. The mass and size differential power law indices, $s$ and $\gamma$, are related by the following equations: $s = (\gamma+2)/3$, $\gamma = 3s-2$. }
\end{tablenotes}
%

\end{table*}


\section{Discussion}

The presence of dark matter in the Universe is widely accepted, supported by evidence such as the flat rotation curves of spiral galaxies, the galaxy orbits, hot gas in clusters, and gravitational lensing of background galaxies \citep{Freese2009, Courteau2014}. A fundamental question, however, is to what extent this dark matter is baryonic or non-baryonic and how it influences the galaxy formation and evolution. 

For example, the 'missing mass problem' observed in rotation curves might partially stem from the currently low estimates of dust mass in galaxies, which is often assumed to be about 1\% of the gas mass. However, these estimates are unreliable, as current dust models limit the estimated dust mass in galaxies to align with Big Bang (BB) theory and the $\Lambda$CDM model, assuming the absence of large dust grains and particles in the ISM. This leads to a 'circularity problem' where studies trace only a small fraction of baryonic dark matter, specifically submicron-sized dust grains coupled with gas. This dust component is detectable by reddening and significantly affects the extinction curve. In contrast, larger dust grains interact weakly with light, making them difficult to detect using standard methods based on absorption/emission of dust. As noted by Galliano et al. \citep{Galliano2011}, derived dust mass can be regarded as an \textit{a priori} constraint rather than a genuine result. If we acknowledge the presence of large particles and bodies in the ISM, observations would support a galaxy model in which the mass of baryonic dark matter is at least one to two orders of magnitude greater than currently assumed. 

There are several key pieces of evidence that baryonic dark matter in galaxies is significantly more abundant than current estimates:

\textit{Size distribution of asteroids and planets.} Observations of large particles and compact bodies across various environments confirm a power-law size distribution $dN \sim a^{-\gamma} da$ with an index $\gamma = 3.5-4.0$, valid over a wide size range from micrometres to hundreds of thousands kilometres (see Figure~\ref{fig:4} and Table~\ref{Table:2}). This distribution includes large dust grains and meteoroids detected by satellites such as Ulysses, Copernicus, Cassini and others \citep{Frisch1999, Grun2001, Altobelli2005, Altobelli2007, Kruger_Grun2009, Kruger2015}, near-Earth objects colliding with Earth \citep{Halliday1989, Ceplecha1998, Brown2002, Bland_Artemieva2006}, and fragments in the Main Asteroid Belt and the Kuiper Belt \citep{Bottke2005a, Gladman2009, Schlichting2013, Fuentes2010}. It also applies to exoplanets detected and confirmed by the Kepler and CoRoT space telescopes \citep{Ananyeva2020}. The broad validity of this power-law size distribution with an index $\gamma = 3.5-4.0$ is further supported  by laboratory measurements and collision dynamics models on dust destruction \citep{Deckers_Teiser2014}. Since interstellar disks interact with the ISM and a considerable portion of mass of planetary systems is ejected into interstellar space, this power-law dust size distribution likely applies to interstellar dust as well.

\textit{Interstellar object 'Oumuamua.} The discovery of the interstellar object 'Oumuamua, observed by the Pan-STARRS1 telescope on October 19, 2017, and identified by its hyperbolic orbit \citep{Meech2017, Bolin2018, Trilling2018}, suggests a higher density of such objects than previously assumed \citep{Do2018, Portegies-Zwart2018}. These density estimates align with the predictions of the CB model proposed in this paper (see Figure~\ref{fig:2}b,d).

\textit{Submm excess.} The widespread detection of the FIR and (sub)millimetre excess in galaxy emission spectra (see Figure~\ref{fig:3}) suggests the presence of large, cold particles in the ISM. This excess has been discussed extensively by various authors \citep{Lisenfeld2002, Dumke2004, Galliano2003, Galliano2005, Galametz2009, Galametz2011, Paradis2009}, though it has often been dismissed due to concerns that the resulting dust-to-gas ratio would be unrealistic \citep{Bendo2006, Gordon2014, Hermelo2016} or doubts regarding the formation of such bodies \citep{Whitworth2019}. These arguments are problematic because they rely on dust models that impose an unrealistic maximum size limit on dust grains, restricting them to be only a few microns. Ignoring large, cold particles can lead to a significant underestimation of both the dust mass and the dust-to-gas ratio, potentially by  one to two orders of  magnitude or even more.

\textit{	Black holes and other extremely massive cold compact objects.}
For completeness, it should be noted that a portion of dark baryonic matter is contained in black holes (BHs) \citep{Ferrarese_Ford_2005, Volonteri2012, Greene2020}. Black holes exist in three mass categories: stellar-mass black holes ($M \sim 3-20 \, M_\odot$), intermediate-mass black holes ($M \sim 10^2-10^4 \, M_\odot$), and supermassive black holes ($M \sim 10^6-10^{10} \, M_\odot$) typically located at the centres of galaxies \citep{Miller_Colbert_2004, Kormendy_Ho_2013}. BHs are often identified through X-ray emissions from accretion processes, though isolated stellar-mass BHs, the number of which is estimated to be an order of $10^8$ in our Galaxy \citep{Tsuna_Kawanaka_2019}, are radiatively inefficient, making them difficult to observe. 
Additionally, the masses of two of the most well-studied supermassive BHs provide insight into the substantial role BHS play in contributing to baryonic dark matter in galaxies. These include the supermassive BH at the centre of our Galaxy, with $M \approx 3.7 \times 10^6 \, M_\odot$ \citep{Schodel2002, Ghez2005}, and the supermassive BH in galaxy M87, with $M \approx 6.6 \times 10^9 \, M_\odot$ \citep{Gebhardt2011}. Note that formation processes of supermassive BHs are still not fully understood, and observations of supermassive BHs in the early Universe \citep{Banados2018, Dolgov2018, Larson2023, Goulding2023} present challenges to the standard cosmological model and the BB theory.

\section{Conclusion}

Current observations suggest  that the Zwicky's original idea of substantial mass of baryonic dark matter in galaxies and galaxy clusters \citep{Zwicky1937, Zwicky1933} was prematurely rejected without solid evidence. The rejection aimed to support the BBN and BAO theories and justify the $\Lambda$CDM model. However, the current galaxy dust models  \citep{Mathis1977, Draine_Lee1984, Weingartner_Draine2001, Draine_Li2007}, designed to align with the $\Lambda$CDM model face fundamental challenges and inconsistencies with observations. Specifically, these models suffer from a circularity problem, as they neglect the absorption/emission effects of large dust grains and compact bodies, considering only grains smaller than a few microns. Observations of other baryonic dark matter components, such as supermassive black holes in the early universe \citep{Banados2018, Dolgov2018, Larson2023, Goulding2023}, deepen the tensions with the current baryonic dark matter models. 

In the proposed CB model, the restriction to small grains is removed. Based on observations, the dust component of baryonic dark matter is assumed to consist of grains, particles, fragments, and solid bodies that follow a power-law size distribution with a size index between 3.5 and 3.75, slightly varying over the size range from 10$^{-9}$ to 10$^{8}$ m. In the CB model, most of the mass is concentrated in large compact bodies located in galaxy disks, in contrast to non-baryonic dark matter, which is assumed to reside in galaxy halos. The baryonic dark matter need not be gravitationally bound to stars or coupled with gas, except for submicron dust grains, which often form part of gas clouds. Large dust grains are presumed to form efficiently from supernovae ejecta and other stellar flows, producing substantial amounts of macroscopic particles and solid bodies not only in circumstellar disks but also throughout the ISM.

Consequently, baryonic dark matter may be ubiquitous throughout the galaxy, with most of its mass concentrated in large bodies. These bodies are challenging to detect using standard galaxy SED analysis, and their number density is considerably lower than that of planetesimals in circumstellar disks. Despite this, the density remains sufficient for interstellar bodies to significantly contribute to the total galactic mass, potentially exceeding current estimates by more than one to two orders of magnitude. Due to their large size, these bodies do not cause reddening and have minimal light extinction because their low number density and small cross-sectional area make their impact negligible compared to small dust grains. Larger bodies interact weakly with light: thus, weak light interaction is a common attribute of both large, cold bodies (including non-accreting BHs) and  hypothesized non-baryonic dark matter. Notably, a significantly higher amount of baryonic dark matter in the Universe would challenge current estimates of the age of the Universe and question the validity of the standard cosmological model.

\section{Declaration of competing interest}
The author declares to have no competing financial interests.

\section{Data availability}
No new data were analyzed in this paper.




\begin{thebibliography}{100}
\expandafter\ifx\csname url\endcsname\relax
  \def\url#1{\texttt{#1}}\fi
\expandafter\ifx\csname urlprefix\endcsname\relax\def\urlprefix{URL }\fi
\expandafter\ifx\csname href\endcsname\relax
  \def\href#1#2{#2} \def\path#1{#1}\fi

\bibitem{Zwicky1933}
F.~{Zwicky}, {Republication of: The redshift of extragalactic nebulae}, General
  Relativity and Gravitation 41~(1) (2009) 207--224.


\bibitem{Zwicky1937}
F.~{Zwicky}, {On the Masses of Nebulae and of Clusters of Nebulae},  Astrophysical Journal 86
  (1937) 217.


\bibitem{Rubin_Ford1970}
V.~C. {Rubin}, W.~K. {Ford}, {Rotation of the Andromeda Nebula from a
  Spectroscopic Survey of Emission Regions},  Astrophysical Journal 159 (1970) 379.


\bibitem{Rubin1980}
V.~C. {Rubin}, W.~K. {Ford}, N.~{Thonnard}, {Rotational properties of 21 SC
  galaxies with a large range of luminosities and radii, from NGC 4605 (R=4kpc)
  to UGC 2885 (R=122kpc).},  Astrophysical Journal 238 (1980) 471--487.

\bibitem{Rubin1985}
V.~C. {Rubin}, D.~{Burstein}, W.~K. {Ford}, N.~{Thonnard}, {Rotation
  velocities of 16 SA galaxies and a comparison of Sa, SB and SC rotation
  properties.},  Astrophysical Journal 289 (1985) 81--104.

\bibitem{Albada1985}
T.~S. {van Albada}, J.~N. {Bahcall}, K.~{Begeman}, R.~{Sancisi}, {Distribution
  of dark matter in the spiral galaxy NGC 3198.},  Astrophysical Journal 295 (1985) 305--313.

\bibitem{Kent1987}
S.~M. {Kent}, {Dark Matter in Spiral Galaxies. II. Galaxies with H I Rotation
  Curves},  Astronomical Journal 93 (1987) 816.

\bibitem{Begeman1989}
K.~G. {Begeman}, {HI rotation curves of spiral galaxies. I. NGC 3198.}, Astronomy and Astrophysics
  223 (1989) 47--60.

\bibitem{Persic1996}
M.~{Persic}, P.~{Salucci}, F.~{Stel}, {The universal rotation curve of spiral
  galaxies {\textemdash} I. The dark matter connection}, Monthly Notices of the Royal Astronomical Society 281~(1) (1996)
  27--47.


\bibitem{Sanders1996}
R.~H. {Sanders}, {The Published Extended Rotation Curves of Spiral Galaxies:
  Confrontation with Modified Dynamics},  Astrophysical Journal 473 (1996) 117.


\bibitem{Sofue_Rubin2001}
Y.~{Sofue}, V.~{Rubin}, {Rotation Curves of Spiral Galaxies}, Annual Review of Astronomy and Astrophysics  39 (2001)
  137--174.

\bibitem{deBlok2002}
W.~J.~G. {de Blok}, A.~{Bosma}, {High-resolution rotation curves of low surface
  brightness galaxies}, Astronomy and Astrophysics 385 (2002) 816--846.


\bibitem{Dubinski_Carlberg1991}
J.~{Dubinski}, R.~G. {Carlberg}, {The Structure of Cold Dark Matter Halos},
   Astrophysical Journal 378 (1991) 496.

\bibitem{Navarro1996}
J.~F. {Navarro}, C.~S. {Frenk}, S.~D.~M. {White}, {The Structure of Cold Dark
  Matter Halos},  Astrophysical Journal 462 (1996) 563.


\bibitem{Navarro1997}
J.~F. {Navarro}, C.~S. {Frenk}, S.~D.~M. {White}, {A Universal Density Profile
  from Hierarchical Clustering},  Astrophysical Journal 490~(2) (1997) 493--508.


\bibitem{White_Rees1978}
S.~D.~M. {White}, M.~J. {Rees}, {Core condensation in heavy halos: a two-stage
  theory for galaxy formation and clustering}, Monthly Notices of the Royal Astronomical Society 183 (1978) 341--358.

\bibitem{Davis1985}
M.~{Davis}, G.~{Efstathiou}, C.~S. {Frenk}, S.~D.~M. {White}, {The evolution of
  large-scale structure in a universe dominated by cold dark matter},  Astrophysical Journal 292
  (1985) 371--394.

\bibitem{White1987}
S.~D.~M. {White}, C.~S. {Frenk}, M.~{Davis}, G.~{Efstathiou}, {Clusters,
  Filaments, and Voids in a Universe Dominated by Cold Dark Matter},  Astrophysical Journal 313
  (1987) 505.


\bibitem{Maddox1990a}
S.~J. {Maddox}, G.~{Efstathiou}, W.~J. {Sutherland}, J.~{Loveday}, {Galaxy
  correlations on large scales.}, Monthly Notices of the Royal Astronomical Society 242 (1990) 43.

\bibitem{Moore1999}
B.~{Moore}, S.~{Ghigna}, F.~{Governato}, G.~{Lake}, T.~{Quinn}, J.~{Stadel},
  P.~{Tozzi}, {Dark Matter Substructure within Galactic Halos},  Astrophysical Journal 524~(1)
  (1999) L19--L22.

\bibitem{Bergstrom2000}
L.~{Bergstr{\"o}m}, {Non-baryonic dark matter: observational evidence and
  detection methods}, Reports on Progress in Physics 63~(5) (2000) 793--841.


\bibitem{DelPopolo2013}
A.~{Del Popolo}, {Non-baryonic dark matter in cosmology}, in: L.~A.
  {Urena-L{\'o}pez}, R.~{Becerril-B{\'a}rcenas}, R.~{Linares-Romero} (Eds.), IX
  Mexican School on Gravitation and Mathematical Physics: Cosmology for the
  XXIst Century: Gravitation and Mathematical Physics Division of the Mexican
  Physical Society DGFM-SMF, Vol. 1548 of American Institute of Physics
  Conference Series, 2013, pp. 2--63.


\bibitem{Bertone_Hooper2018}
G.~{Bertone}, D.~{Hooper}, {History of dark matter}, Reviews of Modern Physics
  90~(4) (2018) 045002.


\bibitem{Kroupa2012}
P.~{Kroupa}, {The Dark Matter Crisis: Falsification of the Current Standard
  Model of Cosmology}, Publications of the Astronomical Society of Australia 29
  (2012) 395--433.

\bibitem{Kroupa2015}
P.~{Kroupa}, {Galaxies as simple dynamical systems: observational data disfavor
  dark matter and stochastic star formation}, Canadian Journal of Physics 93
  (2015) 169--202.

\bibitem{Del-Popolo_Le-Delliou2017}
A.~{Del Popolo}, M.~{Le Delliou}, {Small Scale Problems of the
  {\ensuremath{\Lambda}}CDM Model: A Short Review}, Galaxies 5~(1) (2017) 17.

\bibitem{Milgrom1983a}
M.~{Milgrom}, {A modification of the Newtonian dynamics as a possible
  alternative to the hidden mass hypothesis.},  Astrophysical Journal 270 (1983) 365--370.


\bibitem{Milgrom2012}
M.~{Milgrom}, {Testing MOND over a Wide Acceleration Range in X-Ray
  Ellipticals}, Physical Review Letters 109~(13) (2012) 131101.

\bibitem{Bekenstein2004}
J.~D. {Bekenstein}, {Relativistic gravitation theory for the modified Newtonian
  dynamics paradigm}, Physical Review D 70~(8) (2004) 083509.


\bibitem{Mannheim2012}
P.~D. {Mannheim}, {Making the Case for Conformal Gravity}, Foundations of
  Physics 42~(3) (2012) 388--420.


\bibitem{Mannheim2019}
P.~D. {Mannheim}, {Is dark matter fact or fantasy? {\textemdash} Clues from the
  data}, International Journal of Modern Physics D 28~(14) (2019) 1944022.


\bibitem{Vavrycuk_Frontiers_Astron_Space_Sci_2023}
V.~{Vavry{\v{c}}uk},
  {Gravitational
  orbits in the expanding Universe revisited}, Frontiers in Astronomy and
  Space Sciences 10 (2023) 1071743.


\bibitem{Galliano2005}
F.~{Galliano}, S.~C. {Madden}, A.~P. {Jones}, C.~D. {Wilson}, J.~P. {Bernard},
  {ISM properties in low-metallicity environments. III. The spectral energy
  distributions of II Zw 40, He 2-10 and NGC 1140}, Astronomy and Astrophysics 434~(3) (2005)
  867--885.

\bibitem{Ade2011_Planck_XVII_Submm_excess}
{Planck Collaboration}, P.~A.~R. {Ade}, N.~{Aghanim}, M.~{Arnaud},
  M.~{Ashdown}, J.~{Aumont}, C.~{Baccigalupi}, A.~{Balbi}, A.~J. {Banday},
  R.~B. {Barreiro}, J.~G. {Bartlett}, E.~{Battaner}, K.~{Benabed},
  A.~{Beno{\^\i}t}, J.~P. {Bernard}, M.~{Bersanelli}, R.~{Bhatia}, J.~J.
  {Bock}, A.~{Bonaldi}, J.~R. {Bond}, J.~{Borrill}, C.~{Bot}, F.~R. {Bouchet},
  F.~{Boulanger}, M.~{Bucher}, C.~{Burigana}, P.~{Cabella}, J.~F. {Cardoso},
  A.~{Catalano}, L.~{Cay{\'o}n}, A.~{Challinor}, A.~{Chamballu}, L.~Y.
  {Chiang}, C.~{Chiang}, P.~R. {Christensen}, D.~L. {Clements}, S.~{Colombi},
  F.~{Couchot}, A.~{Coulais}, B.~P. {Crill}, F.~{Cuttaia}, L.~{Danese}, R.~D.
  {Davies}, R.~J. {Davis}, P.~{de Bernardis}, G.~{de Gasperis}, A.~{de Rosa},
  G.~{de Zotti}, J.~{Delabrouille}, J.~M. {Delouis}, F.~X. {D{\'e}sert},
  C.~{Dickinson}, K.~{Dobashi}, S.~{Donzelli}, O.~{Dor{\'e}}, U.~{D{\"o}rl},
  M.~{Douspis}, X.~{Dupac}, G.~{Efstathiou}, T.~A. {En{\ss}lin}, F.~{Finelli},
  O.~{Forni}, M.~{Frailis}, E.~{Franceschi}, Y.~{Fukui}, S.~{Galeotta},
  K.~{Ganga}, M.~{Giard}, G.~{Giardino}, Y.~{Giraud-H{\'e}raud},
  J.~{Gonz{\'a}lez-Nuevo}, K.~M. {G{\'o}rski}, S.~{Gratton}, A.~{Gregorio},
  A.~{Gruppuso}, D.~{Harrison}, G.~{Helou}, S.~{Henrot-Versill{\'e}},
  D.~{Herranz}, S.~R. {Hildebrandt}, E.~{Hivon}, M.~{Hobson}, W.~A. {Holmes},
  W.~{Hovest}, R.~J. {Hoyland}, K.~M. {Huffenberger}, A.~H. {Jaffe}, W.~C.
  {Jones}, M.~{Juvela}, A.~{Kawamura}, E.~{Keih{\"a}nen}, R.~{Keskitalo}, T.~S.
  {Kisner}, R.~{Kneissl}, L.~{Knox}, H.~{Kurki-Suonio}, G.~{Lagache},
  A.~{L{\"a}hteenm{\"a}ki}, J.~M. {Lamarre}, A.~{Lasenby}, R.~J. {Laureijs},
  C.~R. {Lawrence}, S.~{Leach}, R.~{Leonardi}, C.~{Leroy},
  M.~{Linden-V{\o}rnle}, M.~{L{\'o}pez-Caniego}, P.~M. {Lubin}, J.~F.
  {Mac{\'\i}as-P{\'e}rez}, C.~J. {MacTavish}, S.~{Madden}, B.~{Maffei},
  N.~{Mandolesi}, R.~{Mann}, M.~{Maris}, E.~{Mart{\'\i}nez-Gonz{\'a}lez},
  S.~{Masi}, S.~{Matarrese}, F.~{Matthai}, P.~{Mazzotta}, P.~R. {Meinhold},
  A.~{Melchiorri}, L.~{Mendes}, A.~{Mennella}, M.~A. {Miville-Desch{\^e}nes},
  A.~{Moneti}, L.~{Montier}, G.~{Morgante}, D.~{Mortlock}, D.~{Munshi},
  A.~{Murphy}, P.~{Naselsky}, F.~{Nati}, P.~{Natoli}, C.~B. {Netterfield},
  H.~U. {N{\o}rgaard-Nielsen}, F.~{Noviello}, D.~{Novikov}, I.~{Novikov},
  T.~{Onishi}, S.~{Osborne}, F.~{Pajot}, R.~{Paladini}, D.~{Paradis},
  F.~{Pasian}, G.~{Patanchon}, O.~{Perdereau}, L.~{Perotto}, F.~{Perrotta},
  F.~{Piacentini}, M.~{Piat}, S.~{Plaszczynski}, E.~{Pointecouteau},
  G.~{Polenta}, N.~{Ponthieu}, T.~{Poutanen}, G.~{Pr{\'e}zeau}, S.~{Prunet},
  J.~L. {Puget}, W.~T. {Reach}, R.~{Rebolo}, M.~{Reinecke}, C.~{Renault},
  S.~{Ricciardi}, T.~{Riller}, I.~{Ristorcelli}, G.~{Rocha}, C.~{Rosset},
  M.~{Rowan-Robinson}, J.~A. {Rubi{\~n}o-Mart{\'\i}n}, B.~{Rusholme},
  M.~{Sandri}, G.~{Savini}, D.~{Scott}, M.~D. {Seiffert}, G.~F. {Smoot}, J.~L.
  {Starck}, F.~{Stivoli}, V.~{Stolyarov}, R.~{Sudiwala}, J.~F. {Sygnet}, J.~A.
  {Tauber}, L.~{Terenzi}, L.~{Toffolatti}, M.~{Tomasi}, J.~P. {Torre},
  M.~{Tristram}, J.~{Tuovinen}, G.~{Umana}, L.~{Valenziano}, J.~{Varis},
  P.~{Vielva}, F.~{Villa}, N.~{Vittorio}, L.~A. {Wade}, B.~D. {Wandelt},
  A.~{Wilkinson}, N.~{Ysard}, D.~{Yvon}, A.~{Zacchei}, A.~{Zonca}, {Planck
  early results. XVII. Origin of the submillimetre excess dust emission in the
  Magellanic Clouds}, Astronomy and Astrophysics 536 (2011) A17.


\bibitem{Galametz2011}
M.~{Galametz}, S.~C. {Madden}, F.~{Galliano}, S.~{Hony}, G.~J. {Bendo},
  M.~{Sauvage}, {Probing the dust properties of galaxies up to submillimetre
  wavelengths. II. Dust-to-gas mass ratio trends with metallicity and the submm
  excess in dwarf galaxies}, Astronomy and Astrophysics 532 (2011) A56.


\bibitem{Dale2012}
D.~A. {Dale}, G.~{Aniano}, C.~W. {Engelbracht}, J.~L. {Hinz}, O.~{Krause},
  E.~J. {Montiel}, H.~{Roussel}, P.~N. {Appleton}, L.~{Armus}, P.~{Beir{\~a}o},
  A.~D. {Bolatto}, B.~R. {Brandl}, D.~{Calzetti}, A.~F. {Crocker}, K.~V.
  {Croxall}, B.~T. {Draine}, M.~{Galametz}, K.~D. {Gordon}, B.~A. {Groves},
  C.~N. {Hao}, G.~{Helou}, L.~K. {Hunt}, B.~D. {Johnson}, R.~C. {Kennicutt},
  J.~{Koda}, A.~K. {Leroy}, Y.~{Li}, S.~E. {Meidt}, A.~E. {Miller}, E.~J.
  {Murphy}, N.~{Rahman}, H.~W. {Rix}, K.~M. {Sandstrom}, M.~{Sauvage},
  E.~{Schinnerer}, R.~A. {Skibba}, J.~D.~T. {Smith}, F.~S. {Tabatabaei},
  F.~{Walter}, C.~D. {Wilson}, M.~G. {Wolfire}, S.~{Zibetti}, {Herschel
  Far-infrared and Submillimeter Photometry for the KINGFISH Sample of nearby
  Galaxies},  Astrophysical Journal 745~(1) (2012) 95.


\bibitem{Remy-Ruyer2013}
A.~{R{\'e}my-Ruyer}, S.~C. {Madden}, F.~{Galliano}, S.~{Hony}, M.~{Sauvage},
  G.~J. {Bendo}, H.~{Roussel}, M.~{Pohlen}, M.~W.~L. {Smith}, M.~{Galametz},
  D.~{Cormier}, V.~{Lebouteiller}, R.~{Wu}, M.~{Baes}, M.~J. {Barlow},
  M.~{Boquien}, A.~{Boselli}, L.~{Ciesla}, I.~{De Looze}, O.~{\L}.
  {Karczewski}, P.~{Panuzzo}, L.~{Spinoglio}, M.~{Vaccari}, C.~D. {Wilson},
  {Revealing the cold dust in low-metallicity environments. I. Photometry
  analysis of the Dwarf Galaxy Survey with Herschel}, Astronomy and Astrophysics 557 (2013) A95.


\bibitem{Hermelo2016}
I.~{Hermelo}, M.~{Rela{\~n}o}, U.~{Lisenfeld}, S.~{Verley}, C.~{Kramer},
  T.~{Ruiz-Lara}, M.~{Boquien}, E.~M. {Xilouris}, M.~{Albrecht}, {Millimeter
  and submillimeter excess emission in M 33 revealed by Planck and LABOCA},
  Astronomy and Astrophysics 590 (2016) A56.


\bibitem{Dale2017}
D.~A. {Dale}, D.~O. {Cook}, H.~{Roussel}, J.~A. {Turner}, L.~{Armus}, A.~D.
  {Bolatto}, M.~{Boquien}, M.~J.~I. {Brown}, D.~{Calzetti}, I.~{De Looze},
  M.~{Galametz}, K.~D. {Gordon}, B.~A. {Groves}, T.~H. {Jarrett}, G.~{Helou},
  R.~{Herrera-Camus}, J.~L. {Hinz}, L.~K. {Hunt}, R.~C. {Kennicutt}, E.~J.
  {Murphy}, A.~{Rest}, K.~M. {Sandstrom}, J.~D.~T. {Smith}, F.~S. {Tabatabaei},
  C.~D. {Wilson}, {Updated 34-band Photometry for the Sings/KINGFISH Samples of
  Nearby Galaxies},  Astrophysical Journal 837~(1) (2017) 90.


\bibitem{Turner2019}
J.~A. {Turner}, D.~A. {Dale}, A.~{Adamo}, D.~{Calzetti}, K.~{Grasha}, E.~K.
  {Grebel}, K.~E. {Johnson}, J.~C. {Lee}, L.~J. {Smith}, I.~{Yoon}, {An
  ALMA/HST Study of Millimeter Dust Emission and Star Clusters},  Astrophysical Journal 884~(2)
  (2019) 112.


\bibitem{Grun2001}
E.~{Gr{\"u}n}, H.~{Kr{\"u}ger}, M.~{Landgraf}, {Cosmic dust}, 2001, pp.
  373--404.

\bibitem{Strub2015}
P.~{Strub}, H.~{Kr{\"u}ger}, V.~J. {Sterken}, {Sixteen Years of Ulysses
  Interstellar Dust Measurements in the Solar System. II. Fluctuations in the
  Dust Flow from the Data},  Astrophysical Journal 812~(2) (2015) 140.


\bibitem{Ivezic2001}
{\v{Z}}.~{Ivezi{\'c}}, S.~{Tabachnik}, R.~{Rafikov}, R.~H. {Lupton},
  T.~{Quinn}, M.~{Hammergren}, L.~{Eyer}, J.~{Chu}, J.~C. {Armstrong},
  X.~{Fan}, K.~{Finlator}, T.~R. {Geballe}, J.~E. {Gunn}, G.~S. {Hennessy},
  G.~R. {Knapp}, S.~K. {Leggett}, J.~A. {Munn}, J.~R. {Pier}, C.~M. {Rockosi},
  D.~P. {Schneider}, M.~A. {Strauss}, B.~{Yanny}, J.~{Brinkmann}, I.~{Csabai},
  R.~B. {Hindsley}, S.~{Kent}, D.~Q. {Lamb}, B.~{Margon}, T.~A. {McKay}, J.~A.
  {Smith}, P.~{Waddel}, D.~G. {York}, {SDSS Collaboration}, {Solar System
  Objects Observed in the Sloan Digital Sky Survey Commissioning Data},  Astronomical Journal
  122~(5) (2001) 2749--2784.


\bibitem{Jedicke2002}
R.~{Jedicke}, J.~{Larsen}, T.~{Spahr}, {Observational Selection Effects in
  Asteroid Surveys}, 2002, pp. 71--87.

\bibitem{Bottke2005a}
W.~F. {Bottke}, D.~D. {Durda}, D.~{Nesvorn{\'y}}, R.~{Jedicke},
  A.~{Morbidelli}, D.~{Vokrouhlick{\'y}}, H.~{Levison}, {The fossilized size
  distribution of the main asteroid belt}, Icarus 175~(1) (2005) 111--140.

\bibitem{Gladman2009}
B.~J. {Gladman}, D.~R. {Davis}, C.~{Neese}, R.~{Jedicke}, G.~{Williams}, J.~J.
  {Kavelaars}, J.-M. {Petit}, H.~{Scholl}, M.~{Holman}, B.~{Warrington},
  G.~{Esquerdo}, P.~{Tricarico}, {On the asteroid belt's orbital and size
  distribution}, Icarus 202~(1) (2009) 104--118.

\bibitem{Mathis1977}
J.~S. {Mathis}, W.~{Rumpl}, K.~H. {Nordsieck}, {The size distribution of
  interstellar grains.},  Astrophysical Journal 217 (1977) 425--433.

\bibitem{Draine_Lee1984}
B.~T. {Draine}, H.~M. {Lee}, {Optical Properties of Interstellar Graphite and
  Silicate Grains},  Astrophysical Journal 285 (1984) 89.

\bibitem{Kim1994}
S.-H. {Kim}, P.~G. {Martin}, P.~D. {Hendry}, {The Size Distribution of
  Interstellar Dust Particles as Determined from Extinction},  Astrophysical Journal 422 (1994)
  164.

\bibitem{Weingartner_Draine2001}
J.~C. {Weingartner}, B.~T. {Draine}, {Dust Grain-Size Distributions and
  Extinction in the Milky Way, Large Magellanic Cloud, and Small Magellanic
  Cloud},  Astrophysical Journal 548 (2001) 296--309.

\bibitem{Draine2003b}
B.~T. {Draine}, {Scattering by Interstellar Dust Grains. II. X-Rays},  Astrophysical Journal
  598~(2) (2003) 1026--1037.

\bibitem{Zubko2004}
V.~{Zubko}, E.~{Dwek}, R.~G. {Arendt}, {Interstellar Dust Models Consistent
  with Extinction, Emission, and Abundance Constraints},  Astrophysical Journals 152~(2) (2004)
  211--249.

\bibitem{Draine2003}
B.~T. {Draine}, {Interstellar Dust Grains}, Annual Review of Astronomy and Astrophysics  41 (2003) 241--289.

\bibitem{Draine2011}
B.~T. {Draine}, {Physics of the Interstellar and Intergalactic Medium}, 2011.

\bibitem{Draine_Salpeter1979a}
B.~T. {Draine}, E.~E. {Salpeter}, {Destruction mechanisms for interstellar
  dust.},  Astrophysical Journal 231 (1979) 438--455.

\bibitem{Draine_Salpeter1979b}
B.~T. {Draine}, E.~E. {Salpeter}, {On the physics of dust grains in hot gas.},
   Astrophysical Journal 231 (1979) 77--94.

\bibitem{Dominik_Tielens1997}
C.~{Dominik}, A.~G.~G.~M. {Tielens}, {The Physics of Dust Coagulation and the
  Structure of Dust Aggregates in Space},  Astrophysical Journal 480~(2) (1997) 647--673.

\bibitem{Dwek1998}
E.~{Dwek}, {The Evolution of the Elemental Abundances in the Gas and Dust
  Phases of the Galaxy},  Astrophysical Journal 501 (1998) 643.

\bibitem{Jones1996}
A.~P. {Jones}, A.~G.~G.~M. {Tielens}, D.~J. {Hollenbach}, {Grain Shattering in
  Shocks: The Interstellar Grain Size Distribution},  Astrophysical Journal 469 (1996) 740.

\bibitem{Hirashita_Kuo2011}
H.~{Hirashita}, T.-M. {Kuo}, {Effects of grain size distribution on the
  interstellar dust mass growth}, Monthly Notices of the Royal Astronomical Society 416~(2) (2011) 1340--1353.

\bibitem{McKinnon2016}
R.~{McKinnon}, P.~{Torrey}, M.~{Vogelsberger}, {Dust formation in Milky
  Way-like galaxies}, Monthly Notices of the Royal Astronomical Society 457~(4) (2016) 3775--3800.

\bibitem{Andrews_Williams2005}
S.~M. {Andrews}, J.~P. {Williams}, {Circumstellar Dust Disks in Taurus-Auriga:
  The Submillimeter Perspective},  Astrophysical Journal 631~(2) (2005) 1134--1160.

\bibitem{Rodmann2006}
J.~{Rodmann}, T.~{Henning}, C.~J. {Chandler}, L.~G. {Mundy}, D.~J. {Wilner},
  {Large dust particles in disks around T Tauri stars}, Astronomy and Astrophysics 446~(1) (2006)
  211--221.

\bibitem{van-der-Marel2013}
N.~{van der Marel}, E.~F. {van Dishoeck}, S.~{Bruderer}, T.~{Birnstiel},
  P.~{Pinilla}, C.~P. {Dullemond}, T.~A. {van Kempen}, M.~{Schmalzl}, J.~M.
  {Brown}, G.~J. {Herczeg}, G.~S. {Mathews}, V.~{Geers}, {A Major Asymmetric
  Dust Trap in a Transition Disk}, Science 340~(6137) (2013) 1199--1202.

\bibitem{Kataoka2014}
A.~{Kataoka}, S.~{Okuzumi}, H.~{Tanaka}, H.~{Nomura}, {Opacity of fluffy dust
  aggregates}, Astronomy and Astrophysics 568 (2014) A42.

\bibitem{Weidenschilling2011}
S.~J. {Weidenschilling}, {Initial sizes of planetesimals and accretion of the
  asteroids}, Icarus 214~(2) (2011) 671--684.

\bibitem{Mainzer2011}
A.~{Mainzer}, J.~{Bauer}, T.~{Grav}, J.~{Masiero}, R.~M. {Cutri}, J.~{Dailey},
  P.~{Eisenhardt}, R.~S. {McMillan}, E.~{Wright}, R.~{Walker}, R.~{Jedicke},
  T.~{Spahr}, D.~{Tholen}, R.~{Alles}, R.~{Beck}, H.~{Brand enburg},
  T.~{Conrow}, T.~{Evans}, J.~{Fowler}, T.~{Jarrett}, K.~{Marsh}, F.~{Masci},
  H.~{McCallon}, S.~{Wheelock}, M.~{Wittman}, P.~{Wyatt}, E.~{DeBaun},
  G.~{Elliott}, D.~{Elsbury}, I.~{Gautier}, T., S.~{Gomillion}, D.~{Leisawitz},
  C.~{Maleszewski}, M.~{Micheli}, A.~{Wilkins}, {Preliminary Results from
  NEOWISE: An Enhancement to the Wide-field Infrared Survey Explorer for Solar
  System Science},  Astrophysical Journal 731~(1) (2011) 53.

\bibitem{Tedesco2002}
E.~F. {Tedesco}, P.~V. {Noah}, M.~{Noah}, S.~D. {Price}, {The Supplemental IRAS
  Minor Planet Survey},  Astronomical Journal 123~(2) (2002) 1056--1085.

\bibitem{Brown2018}
{Gaia Collaboration}, A.~G.~A. {Brown}, A.~{Vallenari}, T.~{Prusti}, J.~H.~J.
  {de Bruijne}, C.~{Babusiaux}, C.~A.~L. {Bailer-Jones}, M.~{Biermann}, D.~W.
  {Evans}, L.~{Eyer}, F.~{Jansen}, C.~{Jordi}, S.~A. {Klioner}, U.~{Lammers},
  L.~{Lindegren}, X.~{Luri}, F.~{Mignard}, C.~{Panem}, D.~{Pourbaix},
  S.~{Randich}, P.~{Sartoretti}, H.~I. {Siddiqui}, C.~{Soubiran}, F.~{van
  Leeuwen}, N.~A. {Walton}, F.~{Arenou}, U.~{Bastian}, M.~{Cropper},
  R.~{Drimmel}, D.~{Katz}, M.~G. {Lattanzi}, J.~{Bakker}, C.~{Cacciari},
  J.~{Casta{\~n}eda}, L.~{Chaoul}, N.~{Cheek}, F.~{De Angeli}, C.~{Fabricius},
  R.~{Guerra}, B.~{Holl}, E.~{Masana}, R.~{Messineo}, N.~{Mowlavi},
  K.~{Nienartowicz}, P.~{Panuzzo}, J.~{Portell}, M.~{Riello}, G.~M. {Seabroke},
  P.~{Tanga}, F.~{Th{\'e}venin}, G.~{Gracia-Abril}, G.~{Comoretto},
  M.~{Garcia-Reinaldos}, D.~{Teyssier}, M.~{Altmann}, R.~{Andrae}, M.~{Audard},
  I.~{Bellas-Velidis}, K.~{Benson}, J.~{Berthier}, R.~{Blomme}, P.~{Burgess},
  G.~{Busso}, B.~{Carry}, A.~{Cellino}, G.~{Clementini}, M.~{Clotet},
  O.~{Creevey}, M.~{Davidson}, J.~{De Ridder}, L.~{Delchambre}, A.~{Dell'Oro},
  C.~{Ducourant}, J.~{Fern{\'a}ndez-Hern{\'a}ndez}, M.~{Fouesneau},
  Y.~{Fr{\'e}mat}, L.~{Galluccio}, M.~{Garc{\'\i}a-Torres},
  J.~{Gonz{\'a}lez-N{\'u}{\~n}ez}, J.~J. {Gonz{\'a}lez-Vidal}, E.~{Gosset},
  L.~P. {Guy}, J.~L. {Halbwachs}, N.~C. {Hambly}, D.~L. {Harrison},
  J.~{Hern{\'a}ndez}, D.~{Hestroffer}, S.~T. {Hodgkin}, A.~{Hutton},
  G.~{Jasniewicz}, A.~{Jean-Antoine-Piccolo}, S.~{Jordan}, A.~J. {Korn},
  A.~{Krone-Martins}, A.~C. {Lanzafame}, T.~{Lebzelter}, W.~{L{\"o}ffler},
  M.~{Manteiga}, P.~M. {Marrese}, J.~M. {Mart{\'\i}n-Fleitas}, A.~{Moitinho},
  A.~{Mora}, K.~{Muinonen}, J.~{Osinde}, E.~{Pancino}, T.~{Pauwels}, J.~M.
  {Petit}, A.~{Recio-Blanco}, P.~J. {Richards}, L.~{Rimoldini}, A.~C. {Robin},
  L.~M. {Sarro}, C.~{Siopis}, M.~{Smith}, A.~{Sozzetti}, M.~{S{\"u}veges},
  J.~{Torra}, W.~{van Reeven}, U.~{Abbas}, A.~{Abreu Aramburu}, S.~{Accart},
  C.~{Aerts}, G.~{Altavilla}, M.~A. {{\'A}lvarez}, R.~{Alvarez}, J.~{Alves},
  R.~I. {Anderson}, A.~H. {Andrei}, E.~{Anglada Varela}, E.~{Antiche},
  T.~{Antoja}, B.~{Arcay}, T.~L. {Astraatmadja}, N.~{Bach}, S.~G. {Baker},
  L.~{Balaguer-N{\'u}{\~n}ez}, P.~{Balm}, C.~{Barache}, C.~{Barata},
  D.~{Barbato}, F.~{Barblan}, P.~S. {Barklem}, D.~{Barrado}, M.~{Barros}, M.~A.
  {Barstow}, S.~{Bartholom{\'e} Mu{\~n}oz}, J.~L. {Bassilana}, U.~{Becciani},
  M.~{Bellazzini}, A.~{Berihuete}, S.~{Bertone}, L.~{Bianchi},
  O.~{Bienaym{\'e}}, S.~{Blanco-Cuaresma}, T.~{Boch}, C.~{Boeche},
  A.~{Bombrun}, R.~{Borrachero}, D.~{Bossini}, S.~{Bouquillon}, G.~{Bourda},
  A.~{Bragaglia}, L.~{Bramante}, M.~A. {Breddels}, A.~{Bressan},
  N.~{Brouillet}, T.~{Br{\"u}semeister}, E.~{Brugaletta}, B.~{Bucciarelli},
  A.~{Burlacu}, D.~{Busonero}, A.~G. {Butkevich}, R.~{Buzzi}, E.~{Caffau},
  R.~{Cancelliere}, G.~{Cannizzaro}, T.~{Cantat-Gaudin}, R.~{Carballo},
  T.~{Carlucci}, J.~M. {Carrasco}, L.~{Casamiquela}, M.~{Castellani},
  A.~{Castro-Ginard}, P.~{Charlot}, L.~{Chemin}, A.~{Chiavassa}, G.~{Cocozza},
  G.~{Costigan}, S.~{Cowell}, F.~{Crifo}, M.~{Crosta}, C.~{Crowley},
  J.~{Cuypers}, C.~{Dafonte}, Y.~{Damerdji}, A.~{Dapergolas}, P.~{David},
  M.~{David}, P.~{de Laverny}, F.~{De Luise}, R.~{De March}, D.~{de Martino},
  R.~{de Souza}, A.~{de Torres}, J.~{Debosscher}, E.~{del Pozo}, M.~{Delbo},
  A.~{Delgado}, H.~E. {Delgado}, P.~{Di Matteo}, S.~{Diakite}, C.~{Diener},
  E.~{Distefano}, C.~{Dolding}, P.~{Drazinos}, J.~{Dur{\'a}n}, B.~{Edvardsson},
  H.~{Enke}, K.~{Eriksson}, P.~{Esquej}, G.~{Eynard Bontemps}, C.~{Fabre},
  M.~{Fabrizio}, S.~{Faigler}, A.~J. {Falc{\~a}o}, M.~{Farr{\`a}s Casas},
  L.~{Federici}, G.~{Fedorets}, P.~{Fernique}, F.~{Figueras}, F.~{Filippi},
  K.~{Findeisen}, A.~{Fonti}, E.~{Fraile}, M.~{Fraser}, B.~{Fr{\'e}zouls},
  M.~{Gai}, S.~{Galleti}, D.~{Garabato}, F.~{Garc{\'\i}a-Sedano},
  A.~{Garofalo}, N.~{Garralda}, A.~{Gavel}, P.~{Gavras}, J.~{Gerssen},
  R.~{Geyer}, P.~{Giacobbe}, G.~{Gilmore}, S.~{Girona}, G.~{Giuffrida},
  F.~{Glass}, M.~{Gomes}, M.~{Granvik}, A.~{Gueguen}, A.~{Guerrier},
  J.~{Guiraud}, R.~{Guti{\'e}rrez-S{\'a}nchez}, R.~{Haigron},
  D.~{Hatzidimitriou}, M.~{Hauser}, M.~{Haywood}, U.~{Heiter}, A.~{Helmi},
  J.~{Heu}, T.~{Hilger}, D.~{Hobbs}, W.~{Hofmann}, G.~{Holland}, H.~E.
  {Huckle}, A.~{Hypki}, V.~{Icardi}, K.~{Jan{\ss}en}, G.~{Jevardat de
  Fombelle}, P.~G. {Jonker}, {\'A}.~L. {Juh{\'a}sz}, F.~{Julbe},
  A.~{Karampelas}, A.~{Kewley}, J.~{Klar}, A.~{Kochoska}, R.~{Kohley},
  K.~{Kolenberg}, M.~{Kontizas}, E.~{Kontizas}, S.~E. {Koposov},
  G.~{Kordopatis}, Z.~{Kostrzewa-Rutkowska}, P.~{Koubsky}, S.~{Lambert}, A.~F.
  {Lanza}, Y.~{Lasne}, J.~B. {Lavigne}, Y.~{Le Fustec}, C.~{Le Poncin-Lafitte},
  Y.~{Lebreton}, S.~{Leccia}, N.~{Leclerc}, I.~{Lecoeur-Taibi}, H.~{Lenhardt},
  F.~{Leroux}, S.~{Liao}, E.~{Licata}, H.~E.~P. {Lindstr{\o}m}, T.~A. {Lister},
  E.~{Livanou}, A.~{Lobel}, M.~{L{\'o}pez}, S.~{Managau}, R.~G. {Mann},
  G.~{Mantelet}, O.~{Marchal}, J.~M. {Marchant}, M.~{Marconi}, S.~{Marinoni},
  G.~{Marschalk{\'o}}, D.~J. {Marshall}, M.~{Martino}, G.~{Marton}, N., {Gaia
  Data Release 2. Summary of the contents and survey properties}, Astronomy and Astrophysics 616
  (2018) A1.

\bibitem{Wright1987}
E.~L. {Wright}, {Long-wavelength absorption by fractal dust grains},  Astrophysical Journal 320
  (1987) 818--824.

\bibitem{Henning1995}
T.~{Henning}, B.~{Michel}, R.~{Stognienko}, {Dust opacities in dense regions},
  Planetary Space Science 43 (1995) 1333--1343.

\bibitem{Blum_Wurm2008}
J.~{Blum}, G.~{Wurm}, {The growth mechanisms of macroscopic bodies in
  protoplanetary disks.}, Annual Review of Astronomy and Astrophysics  46 (2008) 21--56.

\bibitem{Chiang_Youdin2010}
E.~{Chiang}, A.~N. {Youdin}, {Forming Planetesimals in Solar and Extrasolar
  Nebulae}, Annual Review of Earth and Planetary Sciences 38 (2010) 493--522.

\bibitem{Ade2015_Planck_XIX_Polarized_emission}
{Planck Collaboration}, P.~A.~R. {Ade}, N.~{Aghanim}, D.~{Alina}, M.~I.~R.
  {Alves}, C.~{Armitage-Caplan}, M.~{Arnaud}, D.~{Arzoumanian}, M.~{Ashdown},
  F.~{Atrio-Barand ela}, J.~{Aumont}, C.~{Baccigalupi}, A.~J. {Banday}, R.~B.
  {Barreiro}, E.~{Battaner}, K.~{Benabed}, A.~{Benoit-L{\'e}vy}, J.~P.
  {Bernard}, M.~{Bersanelli}, P.~{Bielewicz}, J.~J. {Bock}, J.~R. {Bond},
  J.~{Borrill}, F.~R. {Bouchet}, F.~{Boulanger}, A.~{Bracco}, C.~{Burigana},
  R.~C. {Butler}, J.~F. {Cardoso}, A.~{Catalano}, A.~{Chamballu}, R.~R.
  {Chary}, H.~C. {Chiang}, P.~R. {Christensen}, S.~{Colombi}, L.~P.~L.
  {Colombo}, C.~{Combet}, F.~{Couchot}, A.~{Coulais}, B.~P. {Crill},
  A.~{Curto}, F.~{Cuttaia}, L.~{Danese}, R.~D. {Davies}, R.~J. {Davis}, P.~{de
  Bernardis}, E.~M. {de Gouveia Dal Pino}, A.~{de Rosa}, G.~{de Zotti},
  J.~{Delabrouille}, F.~X. {D{\'e}sert}, C.~{Dickinson}, J.~M. {Diego},
  S.~{Donzelli}, O.~{Dor{\'e}}, M.~{Douspis}, J.~{Dunkley}, X.~{Dupac},
  G.~{Efstathiou}, T.~A. {En{\ss}lin}, H.~K. {Eriksen}, E.~{Falgarone},
  K.~{Ferri{\`e}re}, F.~{Finelli}, O.~{Forni}, M.~{Frailis}, A.~A. {Fraisse},
  E.~{Franceschi}, S.~{Galeotta}, K.~{Ganga}, T.~{Ghosh}, M.~{Giard},
  Y.~{Giraud-H{\'e}raud}, J.~{Gonz{\'a}lez-Nuevo}, K.~M. {G{\'o}rski},
  A.~{Gregorio}, A.~{Gruppuso}, V.~{Guillet}, F.~K. {Hansen}, D.~L. {Harrison},
  G.~{Helou}, C.~{Hern{\'a}ndez-Monteagudo}, S.~R. {Hildebrand t}, E.~{Hivon},
  M.~{Hobson}, W.~A. {Holmes}, A.~{Hornstrup}, K.~M. {Huffenberger}, A.~H.
  {Jaffe}, T.~R. {Jaffe}, W.~C. {Jones}, M.~{Juvela}, E.~{Keih{\"a}nen},
  R.~{Keskitalo}, T.~S. {Kisner}, R.~{Kneissl}, J.~{Knoche}, M.~{Kunz},
  H.~{Kurki-Suonio}, G.~{Lagache}, A.~{L{\"a}hteenm{\"a}ki}, J.~M. {Lamarre},
  A.~{Lasenby}, C.~R. {Lawrence}, J.~P. {Leahy}, R.~{Leonardi}, F.~{Levrier},
  M.~{Liguori}, P.~B. {Lilje}, M.~{Linden-V{\o}rnle}, M.~{L{\'o}pez-Caniego},
  P.~M. {Lubin}, J.~F. {Mac{\'\i}as-P{\'e}rez}, B.~{Maffei}, A.~M.
  {Magalh{\~a}es}, D.~{Maino}, N.~{Mandolesi}, M.~{Maris}, D.~J. {Marshall},
  P.~G. {Martin}, E.~{Mart{\'\i}nez-Gonz{\'a}lez}, S.~{Masi}, S.~{Matarrese},
  P.~{Mazzotta}, A.~{Melchiorri}, L.~{Mendes}, A.~{Mennella}, M.~{Migliaccio},
  M.~A. {Miville-Desch{\^e}nes}, A.~{Moneti}, L.~{Montier}, G.~{Morgante},
  D.~{Mortlock}, D.~{Munshi}, J.~A. {Murphy}, P.~{Naselsky}, F.~{Nati},
  P.~{Natoli}, C.~B. {Netterfield}, F.~{Noviello}, D.~{Novikov}, I.~{Novikov},
  C.~A. {Oxborrow}, L.~{Pagano}, F.~{Pajot}, R.~{Paladini}, D.~{Paoletti},
  F.~{Pasian}, T.~J. {Pearson}, O.~{Perdereau}, L.~{Perotto}, F.~{Perrotta},
  F.~{Piacentini}, M.~{Piat}, D.~{Pietrobon}, S.~{Plaszczynski}, F.~{Poidevin},
  E.~{Pointecouteau}, G.~{Polenta}, L.~{Popa}, G.~W. {Pratt}, S.~{Prunet},
  J.~L. {Puget}, J.~P. {Rachen}, W.~T. {Reach}, R.~{Rebolo}, M.~{Reinecke},
  M.~{Remazeilles}, C.~{Renault}, S.~{Ricciardi}, T.~{Riller},
  I.~{Ristorcelli}, G.~{Rocha}, C.~{Rosset}, G.~{Roudier}, J.~A.
  {Rubi{\~n}o-Mart{\'\i}n}, B.~{Rusholme}, M.~{Sandri}, G.~{Savini},
  D.~{Scott}, L.~D. {Spencer}, V.~{Stolyarov}, R.~{Stompor}, R.~{Sudiwala},
  D.~{Sutton}, A.~S. {Suur-Uski}, J.~F. {Sygnet}, J.~A. {Tauber}, L.~{Terenzi},
  L.~{Toffolatti}, M.~{Tomasi}, M.~{Tristram}, M.~{Tucci}, G.~{Umana},
  L.~{Valenziano}, J.~{Valiviita}, B.~{Van Tent}, P.~{Vielva}, F.~{Villa},
  L.~A. {Wade}, B.~D. {Wandelt}, A.~{Zacchei}, A.~{Zonca}, {Planck intermediate
  results. XIX. An overview of the polarized thermal emission from Galactic
  dust}, Astronomy and Astrophysics 576 (2015) A104.

\bibitem{Cardelli1989}
J.~A. {Cardelli}, G.~C. {Clayton}, J.~S. {Mathis}, {The Relationship between
  Infrared, Optical, and Ultraviolet Extinction},  Astrophysical Journal 345 (1989) 245.

\bibitem{Fitzpatrick1999}
E.~L. {Fitzpatrick}, {Correcting for the Effects of Interstellar Extinction},
  Publications of the Astronomical Society of the Pacific 111~(755) (1999) 63--75.

\bibitem{Calzetti2001}
D.~{Calzetti}, {The Dust Opacity of Star-forming Galaxies}, Publications of the Astronomical Society of the Pacific 113 (2001)
  1449--1485.

\bibitem{Beckman1996}
J.~E. {Beckman}, R.~F. {Peletier}, J.~H. {Knapen}, R.~L.~M. {Corradi}, L.~J.
  {Gentet}, {Scale Lengths in Disk Surface Brightness as Probes of Dust
  Extinction in Three Spiral Galaxies: M51, NGC 3631, and NGC 4321},  Astrophysical Journal 467
  (1996) 175.

\bibitem{Holwerda2005a}
B.~W. {Holwerda}, R.~A. {Gonzalez}, R.~J. {Allen}, P.~C. {van der Kruit}, {The
  Opacity of Spiral Galaxy Disks. IV. Radial Extinction Profiles from Counts of
  Distant Galaxies Seen through Foreground Disks},  Astronomical Journal 129 (2005) 1396--1411.

\bibitem{Holwerda2005b}
B.~W. {Holwerda}, R.~A. {Gonzalez}, R.~J. {Allen}, P.~C. {van der Kruit}, {The
  Opacity of Spiral Galaxy Disks. III. Automating the Synthetic Field Method},
   Astronomical Journal 129 (2005) 1381--1395.

\bibitem{Pettini1994}
M.~{Pettini}, L.~J. {Smith}, R.~W. {Hunstead}, D.~L. {King}, {Metal enrichment,
  dust, and star formation in galaxies at high redshifts. 3: Zn and CR
  abundances for 17 damped Lyman-alpha systems},  Astrophysical Journal 426 (1994) 79--96.

\bibitem{Ledoux2002}
C.~{Ledoux}, J.~{Bergeron}, P.~{Petitjean}, {Dust depletion and abundance
  pattern in damped Lyalpha systems: A sample of Mn and Ti abundances at z <
  2.2}, Astronomy and Astrophysics 385 (2002) 802--815.

\bibitem{Watson2015}
D.~{Watson}, L.~{Christensen}, K.~K. {Knudsen}, J.~{Richard}, A.~{Gallazzi},
  M.~J. {Micha{\l}owski}, {A dusty, normal galaxy in the epoch of
  reionization}, Nature 519 (2015) 327--330.

\bibitem{Laporte2017}
N.~{Laporte}, R.~S. {Ellis}, F.~{Boone}, F.~E. {Bauer}, D.~{Qu{\'e}nard}, G.~W.
  {Roberts-Borsani}, R.~{Pell{\'o}}, I.~{P{\'e}rez-Fournon}, A.~{Streblyanska},
  {Dust in the Reionization Era: ALMA Observations of a z=8.38
  Gravitationally Lensed Galaxy},  Astrophysical Journal 837 (2017) L21.

\bibitem{Witt1992}
A.~N. {Witt}, J.~{Thronson}, Harley~A., J.~{Capuano}, John~M., {Dust and the
  Transfer of Stellar Radiation within Galaxies},  Astrophysical Journal 393 (1992) 611.

\bibitem{Silva1998}
L.~{Silva}, G.~L. {Granato}, A.~{Bressan}, L.~{Danese}, {Modeling the Effects
  of Dust on Galactic Spectral Energy Distributions from the Ultraviolet to the
  Millimeter Band},  Astrophysical Journal 509 (1998) 103--117.

\bibitem{Misselt2001}
K.~A. {Misselt}, K.~D. {Gordon}, G.~C. {Clayton}, M.~J. {Wolff}, {The DIRTY
  Model. II. Self-consistent Treatment of Dust Heating and Emission in a
  Three-dimensional Radiative Transfer Code},  Astrophysical Journal 551~(1) (2001) 277--293.

\bibitem{Tuffs2004}
R.~J. {Tuffs}, C.~C. {Popescu}, H.~J. {V{\"o}lk}, N.~D. {Kylafis}, M.~A.
  {Dopita}, {Modelling the spectral energy distribution of galaxies. III.
  Attenuation of stellar light in spiral galaxies}, Astronomy and Astrophysics 419 (2004) 821--835.

\bibitem{Honig2006}
S.~F. {H{\"o}nig}, T.~{Beckert}, K.~{Ohnaka}, G.~{Weigelt}, {Radiative transfer
  modeling of three-dimensional clumpy AGN tori and its application to NGC
  1068}, Astronomy and Astrophysics 452~(2) (2006) 459--471.

\bibitem{Popescu2000}
C.~C. {Popescu}, A.~{Misiriotis}, N.~D. {Kylafis}, R.~J. {Tuffs},
  J.~{Fischera}, {Modelling the spectral energy distribution of galaxies. I.
  Radiation fields and grain heating in the edge-on spiral NGC 891}, Astronomy and Astrophysics 362
  (2000) 138--150.

\bibitem{Popescu2011}
C.~C. {Popescu}, R.~J. {Tuffs}, M.~A. {Dopita}, J.~{Fischera}, N.~D. {Kylafis},
  B.~F. {Madore}, {Modelling the spectral energy distribution of galaxies. V.
  The dust and PAH emission SEDs of disk galaxies}, Astronomy and Astrophysics 527 (2011) A109.

\bibitem{Walcher2011}
J.~{Walcher}, B.~{Groves}, T.~{Budav{\'a}ri}, D.~{Dale}, {Fitting the
  integrated spectral energy distributions of galaxies}, Astrophysics and Space Science 331 (2011)
  1--52.

\bibitem{Cunha2008}
E.~{da Cunha}, S.~{Charlot}, D.~{Elbaz}, {A simple model to interpret the
  ultraviolet, optical and infrared emission from galaxies}, Monthly Notices of the Royal Astronomical Society 388 (2008)
  1595--1617.

\bibitem{Calzetti2000}
D.~{Calzetti}, L.~{Armus}, R.~C. {Bohlin}, A.~L. {Kinney}, J.~{Koornneef},
  T.~{Storchi-Bergmann}, {The Dust Content and Opacity of Actively Star-forming
  Galaxies},  Astrophysical Journal 533 (2000) 682--695.

\bibitem{Galametz2012}
M.~{Galametz}, R.~C. {Kennicutt}, M.~{Albrecht}, G.~{Aniano}, L.~{Armus},
  F.~{Bertoldi}, D.~{Calzetti}, A.~F. {Crocker}, K.~V. {Croxall}, D.~A. {Dale},
  J.~{Donovan Meyer}, B.~T. {Draine}, C.~W. {Engelbracht}, J.~L. {Hinz},
  H.~{Roussel}, R.~A. {Skibba}, F.~S. {Tabatabaei}, F.~{Walter}, A.~{Weiss},
  C.~D. {Wilson}, M.~G. {Wolfire}, {Mapping the cold dust temperatures and
  masses of nearby KINGFISH galaxies with Herschel}, Monthly Notices of the Royal Astronomical Society 425~(1) (2012)
  763--787.

\bibitem{Draine2003a}
B.~T. {Draine}, {Scattering by Interstellar Dust Grains. I. Optical and
  Ultraviolet},  Astrophysical Journal 598~(2) (2003) 1017--1025.

\bibitem{Li_Draine2001b}
A.~{Li}, B.~T. {Draine}, {Infrared Emission from Interstellar Dust. II. The
  Diffuse Interstellar Medium},  Astrophysical Journal 554 (2001) 778--802.

\bibitem{Draine_Li2007}
B.~T. {Draine}, A.~{Li}, {Infrared Emission from Interstellar Dust. IV. The
  Silicate-Graphite-PAH Model in the Post-Spitzer Era},  Astrophysical Journal 657~(2) (2007)
  810--837.

\bibitem{Inoue2020}
A.~K. {Inoue}, T.~{Hashimoto}, H.~{Chihara}, C.~{Koike}, {Radiative equilibrium
  estimates of dust temperature and mass in high-redshift galaxies}, Monthly Notices of the Royal Astronomical Society
  495~(2) (2020) 1577--1592.

\bibitem{Hildebrand1977}
R.~H. {Hildebrand}, S.~E. {Whitcomb}, R.~{Winston}, R.~F. {Stiening}, D.~A.
  {Harper}, S.~H. {Moseley}, {Submillimeter photometry of extragalactic
  objects.},  Astrophysical Journal 216 (1977) 698--705.

\bibitem{Hildebrand1983}
R.~H. {Hildebrand}, {The determination of cloud masses and dust characteristics
  from submillimetre thermal emission.}, Quarterly Journal of the Royal Astronomical Society 24 (1983) 267--282.

\bibitem{Ossenkopf1992}
V.~{Ossenkopf}, T.~{Henning}, J.~S. {Mathis}, {Constraints on cosmic
  silicates.}, Astronomy and Astrophysics 261 (1992) 567--578.

\bibitem{Baes2020}
M.~{Baes}, A.~{Tr{\v{c}}ka}, P.~{Camps}, J.~{Trayford}, A.~{Katsianis},
  L.~{Marchetti}, T.~{Theuns}, M.~{Vaccari}, B.~{Vandenbroucke}, {Infrared
  luminosity functions and dust mass functions in the EAGLE simulation}, Monthly Notices of the Royal Astronomical Society
  494~(2) (2020) 2912--2924.

\bibitem{Dunne2000}
L.~{Dunne}, S.~{Eales}, M.~{Edmunds}, R.~{Ivison}, P.~{Alexander}, D.~L.
  {Clements}, {The SCUBA Local Universe Galaxy Survey - I. First measurements
  of the submillimetre luminosity and dust mass functions}, Monthly Notices of the Royal Astronomical Society 315~(1)
  (2000) 115--139.

\bibitem{Bohlin1978}
R.~C. {Bohlin}, B.~D. {Savage}, J.~F. {Drake}, {A survey of interstellar H I
  from L-alpha absorption measurements. II},  Astrophysical Journal 224 (1978) 132--142.

\bibitem{Draine2007}
B.~T. {Draine}, A.~{Li}, {Infrared Emission from Interstellar Dust. IV. The
  Silicate-Graphite-PAH Model in the Post-Spitzer Era},  Astrophysical Journal 657 (2007)
  810--837.

\bibitem{Munoz-Mateos2009}
J.~C. {Mu{\~n}oz-Mateos}, A.~{Gil de Paz}, S.~{Boissier}, J.~{Zamorano}, D.~A.
  {Dale}, P.~G. {P{\'e}rez-Gonz{\'a}lez}, J.~{Gallego}, B.~F. {Madore},
  G.~{Bendo}, M.~D. {Thornley}, B.~T. {Draine}, A.~{Boselli}, V.~{Buat},
  D.~{Calzetti}, J.~{Moustakas}, J.~{Kennicutt}, R.~C., {Radial Distribution of
  Stars, Gas, and Dust in Sings Galaxies. II. Derived Dust Properties},  Astrophysical Journal
  701~(2) (2009) 1965--1991.

\bibitem{Draine2014}
B.~T. {Draine}, G.~{Aniano}, O.~{Krause}, B.~{Groves}, K.~{Sandstrom},
  R.~{Braun}, A.~{Leroy}, U.~{Klaas}, H.~{Linz}, H.-W. {Rix}, E.~{Schinnerer},
  A.~{Schmiedeke}, F.~{Walter}, {Andromeda's Dust},  Astrophysical Journal 780~(2) (2014) 172.

\bibitem{Aniano2020}
G.~{Aniano}, B.~T. {Draine}, L.~K. {Hunt}, K.~{Sandstrom}, D.~{Calzetti}, R.~C.
  {Kennicutt}, D.~A. {Dale}, M.~{Galametz}, K.~D. {Gordon}, A.~K. {Leroy},
  J.~D.~T. {Smith}, H.~{Roussel}, M.~{Sauvage}, F.~{Walter}, L.~{Armus}, A.~D.
  {Bolatto}, M.~{Boquien}, A.~{Crocker}, I.~{De Looze}, J.~{Donovan Meyer},
  G.~{Helou}, J.~{Hinz}, B.~D. {Johnson}, J.~{Koda}, A.~{Miller}, E.~{Montiel},
  E.~J. {Murphy}, M.~{Rela{\~n}o}, H.~W. {Rix}, E.~{Schinnerer}, R.~{Skibba},
  M.~G. {Wolfire}, C.~W. {Engelbracht}, {Modeling Dust and Starlight in
  Galaxies Observed by Spitzer and Herschel: The KINGFISH Sample},  Astrophysical Journal 889~(2)
  (2020) 150.

\bibitem{Lisenfeld_Ferrara1998}
U.~{Lisenfeld}, A.~{Ferrara}, {Dust-to-Gas Ratio and Metal Abundance in Dwarf
  Galaxies},  Astrophysical Journal 496~(1) (1998) 145--154.

\bibitem{Sandstrom2013}
K.~M. {Sandstrom}, A.~K. {Leroy}, F.~{Walter}, A.~D. {Bolatto}, K.~V.
  {Croxall}, B.~T. {Draine}, C.~D. {Wilson}, M.~{Wolfire}, D.~{Calzetti}, R.~C.
  {Kennicutt}, G.~{Aniano}, J.~{Donovan Meyer}, A.~{Usero}, F.~{Bigiel},
  E.~{Brinks}, W.~J.~G. {de Blok}, A.~{Crocker}, D.~{Dale}, C.~W.
  {Engelbracht}, M.~{Galametz}, B.~{Groves}, L.~K. {Hunt}, J.~{Koda},
  K.~{Kreckel}, H.~{Linz}, S.~{Meidt}, E.~{Pellegrini}, H.~W. {Rix},
  H.~{Roussel}, E.~{Schinnerer}, A.~{Schruba}, K.~F. {Schuster}, R.~{Skibba},
  T.~{van der Laan}, P.~{Appleton}, L.~{Armus}, B.~{Brandl}, K.~{Gordon},
  J.~{Hinz}, O.~{Krause}, E.~{Montiel}, M.~{Sauvage}, A.~{Schmiedeke}, J.~D.~T.
  {Smith}, L.~{Vigroux}, {The CO-to-H$_{2}$ Conversion Factor and Dust-to-gas
  Ratio on Kiloparsec Scales in Nearby Galaxies},  Astrophysical Journal 777~(1) (2013) 5.

\bibitem{Wilner2005}
D.~J. {Wilner}, P.~{D'Alessio}, N.~{Calvet}, M.~J. {Claussen}, L.~{Hartmann},
  {Toward Planetesimals in the Disk around TW Hydrae: 3.5 Centimeter Dust
  Emission},  Astrophysical Journal 626~(2) (2005) L109--L112.

\bibitem{Wyatt2008}
M.~C. {Wyatt}, {Evolution of debris disks.}, Annual Review of Astronomy and Astrophysics  46 (2008) 339--383.

\bibitem{Birnstiel2016}
T.~{Birnstiel}, M.~{Fang}, A.~{Johansen}, {Dust Evolution and the Formation of
  Planetesimals}, Space Science Reviews 205~(1-4) (2016) 41--75.

\bibitem{Portegies-Zwart2018}
S.~{Portegies Zwart}, S.~{Torres}, I.~{Pelupessy}, J.~{B{\'e}dorf}, M.~X.
  {Cai}, {The origin of interstellar asteroidal objects like 1I/2017 U1
  `Oumuamua}, Monthly Notices of the Royal Astronomical Society 479~(1) (2018) L17--L22.

\bibitem{Manser2019}
C.~J. {Manser}, B.~T. {G{\"a}nsicke}, S.~{Eggl}, M.~{Hollands}, P.~{Izquierdo},
  D.~{Koester}, J.~D. {Landstreet}, W.~{Lyra}, T.~R. {Marsh}, F.~{Meru}, A.~J.
  {Mustill}, P.~{Rodr{\'\i}guez-Gil}, O.~{Toloza}, D.~{Veras}, D.~J. {Wilson},
  M.~R. {Burleigh}, M.~B. {Davies}, J.~{Farihi}, N.~{Gentile Fusillo}, D.~{de
  Martino}, S.~G. {Parsons}, A.~{Quirrenbach}, R.~{Raddi}, S.~{Reffert},
  M.~{Del Santo}, M.~R. {Schreiber}, R.~{Silvotti}, S.~{Toonen}, E.~{Villaver},
  M.~{Wyatt}, S.~{Xu}, S.~{Portegies Zwart}, {A planetesimal orbiting within
  the debris disc around a white dwarf star}, Science 364~(6435) (2019) 66--69.

\bibitem{Do2018}
A.~{Do}, M.~A. {Tucker}, J.~{Tonry}, {Interstellar Interlopers: Number Density
  and Origin of {\textquoteleft}Oumuamua-like Objects},  Astrophysical Journal 855~(1) (2018)
  L10.

\bibitem{Whitworth2019}
A.~P. {Whitworth}, K.~A. {Marsh}, P.~J. {Cigan}, J.~J. {Dalcanton}, M.~W.~L.
  {Smith}, H.~L. {Gomez}, O.~{Lomax}, M.~J. {Griffin}, S.~A. {Eales}, {The dust
  in M31}, Monthly Notices of the Royal Astronomical Society 489~(4) (2019) 5436--5452.

\bibitem{Aghanim2016_Planck_XLVIII_Dust_emission}
{Planck Collaboration}, N.~{Aghanim}, M.~{Ashdown}, J.~{Aumont},
  C.~{Baccigalupi}, M.~{Ballardini}, A.~J. {Band ay}, R.~B. {Barreiro},
  N.~{Bartolo}, S.~{Basak}, K.~{Benabed}, J.~P. {Bernard}, M.~{Bersanelli},
  P.~{Bielewicz}, L.~{Bonavera}, J.~R. {Bond}, J.~{Borrill}, F.~R. {Bouchet},
  F.~{Boulanger}, C.~{Burigana}, E.~{Calabrese}, J.~F. {Cardoso}, J.~{Carron},
  H.~C. {Chiang}, L.~P.~L. {Colombo}, B.~{Comis}, F.~{Couchot}, A.~{Coulais},
  B.~P. {Crill}, A.~{Curto}, F.~{Cuttaia}, P.~{de Bernardis}, G.~{de Zotti},
  J.~{Delabrouille}, E.~{Di Valentino}, C.~{Dickinson}, J.~M. {Diego},
  O.~{Dor{\'e}}, M.~{Douspis}, A.~{Ducout}, X.~{Dupac}, S.~{Dusini},
  F.~{Elsner}, T.~A. {En{\ss}lin}, H.~K. {Eriksen}, E.~{Falgarone},
  Y.~{Fantaye}, F.~{Finelli}, F.~{Forastieri}, M.~{Frailis}, A.~A. {Fraisse},
  E.~{Franceschi}, A.~{Frolov}, S.~{Galeotta}, S.~{Galli}, K.~{Ganga}, R.~T.
  {G{\'e}nova-Santos}, M.~{Gerbino}, T.~{Ghosh}, Y.~{Giraud-H{\'e}raud},
  J.~{Gonz{\'a}lez-Nuevo}, K.~M. {G{\'o}rski}, A.~{Gruppuso}, J.~E.
  {Gudmundsson}, F.~K. {Hansen}, G.~{Helou}, S.~{Henrot-Versill{\'e}},
  D.~{Herranz}, E.~{Hivon}, Z.~{Huang}, A.~H. {Jaffe}, W.~C. {Jones},
  E.~{Keih{\"a}nen}, R.~{Keskitalo}, K.~{Kiiveri}, T.~S. {Kisner},
  N.~{Krachmalnicoff}, M.~{Kunz}, H.~{Kurki-Suonio}, J.~M. {Lamarre},
  M.~{Langer}, A.~{Lasenby}, M.~{Lattanzi}, C.~R. {Lawrence}, M.~{Le Jeune},
  F.~{Levrier}, P.~B. {Lilje}, M.~{Lilley}, V.~{Lindholm},
  M.~{L{\'o}pez-Caniego}, Y.~Z. {Ma}, J.~F. {Mac{\'\i}as-P{\'e}rez},
  G.~{Maggio}, D.~{Maino}, N.~{Mand olesi}, A.~{Mangilli}, M.~{Maris}, P.~G.
  {Martin}, E.~{Mart{\'\i}nez-Gonz{\'a}lez}, S.~{Matarrese}, N.~{Mauri}, J.~D.
  {McEwen}, A.~{Melchiorri}, A.~{Mennella}, M.~{Migliaccio}, M.~A.
  {Miville-Desch{\^e}nes}, D.~{Molinari}, A.~{Moneti}, L.~{Montier},
  G.~{Morgante}, A.~{Moss}, P.~{Natoli}, C.~A. {Oxborrow}, L.~{Pagano},
  D.~{Paoletti}, G.~{Patanchon}, O.~{Perdereau}, L.~{Perotto}, V.~{Pettorino},
  F.~{Piacentini}, S.~{Plaszczynski}, L.~{Polastri}, G.~{Polenta}, J.~L.
  {Puget}, J.~P. {Rachen}, B.~{Racine}, M.~{Reinecke}, M.~{Remazeilles},
  A.~{Renzi}, G.~{Rocha}, C.~{Rosset}, M.~{Rossetti}, G.~{Roudier}, J.~A.
  {Rubi{\~n}o-Mart{\'\i}n}, B.~{Ruiz-Granados}, L.~{Salvati}, M.~{Sandri},
  M.~{Savelainen}, D.~{Scott}, C.~{Sirignano}, G.~{Sirri}, J.~D. {Soler}, L.~D.
  {Spencer}, A.~S. {Suur-Uski}, J.~A. {Tauber}, D.~{Tavagnacco}, M.~{Tenti},
  L.~{Toffolatti}, M.~{Tomasi}, M.~{Tristram}, T.~{Trombetti}, J.~{Valiviita},
  F.~{Van Tent}, P.~{Vielva}, F.~{Villa}, N.~{Vittorio}, B.~D. {Wandelt}, I.~K.
  {Wehus}, A.~{Zacchei}, A.~{Zonca}, {Planck intermediate results. XLVIII.
  Disentangling Galactic dust emission and cosmic infrared background
  anisotropies}, Astronomy and Astrophysics 596 (2016) A109.

\bibitem{Gordon2010}
K.~D. {Gordon}, F.~{Galliano}, S.~{Hony}, J.~P. {Bernard}, A.~{Bolatto},
  C.~{Bot}, C.~{Engelbracht}, A.~{Hughes}, F.~P. {Israel}, F.~{Kemper},
  S.~{Kim}, A.~{Li}, S.~C. {Madden}, M.~{Matsuura}, M.~{Meixner}, K.~{Misselt},
  K.~{Okumura}, P.~{Panuzzo}, M.~{Rubio}, W.~T. {Reach}, J.~{Roman-Duval},
  M.~{Sauvage}, R.~{Skibba}, A.~G.~G.~M. {Tielens}, {Determining dust
  temperatures and masses in the Herschel era: The importance of observations
  longward of 200 micron}, Astronomy and Astrophysics 518 (2010) L89.

\bibitem{Tabatabaei2014}
F.~S. {Tabatabaei}, J.~{Braine}, E.~M. {Xilouris}, C.~{Kramer}, M.~{Boquien},
  F.~{Combes}, C.~{Henkel}, M.~{Relano}, S.~{Verley}, P.~{Gratier},
  F.~{Israel}, M.~C. {Wiedner}, M.~{R{\"o}llig}, K.~F. {Schuster}, P.~{van der
  Werf}, {Variation in the dust emissivity index across M 33 with Herschel and
  Spitzer (HerM 33es)}, Astronomy and Astrophysics 561 (2014) A95.

\bibitem{Lisenfeld2002}
U.~{Lisenfeld}, F.~P. {Israel}, J.~M. {Stil}, A.~{Sievers}, {(Sub)millimetre
  emission from from NGC 1569: An abundance of very small grains}, Astronomy and Astrophysics 382
  (2002) 860--871.

\bibitem{Galliano2003}
F.~{Galliano}, S.~C. {Madden}, A.~P. {Jones}, C.~D. {Wilson}, J.~P. {Bernard},
  F.~{Le Peintre}, {ISM properties in low-metallicity environments. II. The
  dust spectral energy distribution of NGC 1569}, Astronomy and Astrophysics 407 (2003) 159--176.

\bibitem{Galametz2009}
M.~{Galametz}, S.~{Madden}, F.~{Galliano}, S.~{Hony}, F.~{Schuller},
  A.~{Beelen}, G.~{Bendo}, M.~{Sauvage}, A.~{Lundgren}, N.~{Billot}, {Probing
  the dust properties of galaxies up to submillimetre wavelengths. I. The
  spectral energy distribution of dwarf galaxies using LABOCA}, Astronomy and Astrophysics 508~(2)
  (2009) 645--664.

\bibitem{Paradis2009}
D.~{Paradis}, J.~P. {Bernard}, C.~{M{\'e}ny}, {Dust emissivity variations in
  the Milky Way}, Astronomy and Astrophysics 506~(2) (2009) 745--756.

\bibitem{Galliano2011}
F.~{Galliano}, S.~{Hony}, J.~P. {Bernard}, C.~{Bot}, S.~C. {Madden},
  J.~{Roman-Duval}, M.~{Galametz}, A.~{Li}, M.~{Meixner}, C.~W. {Engelbracht},
  V.~{Lebouteiller}, K.~{Misselt}, E.~{Montiel}, P.~{Panuzzo}, W.~T. {Reach},
  R.~{Skibba}, {Non-standard grain properties, dark gas reservoir, and extended
  submillimeter excess, probed by Herschel in the Large Magellanic Cloud}, Astronomy and Astrophysics
  536 (2011) A88.

\bibitem{Meixner2010}
M.~{Meixner}, F.~{Galliano}, S.~{Hony}, J.~{Roman-Duval}, T.~{Robitaille},
  P.~{Panuzzo}, M.~{Sauvage}, K.~{Gordon}, C.~{Engelbracht}, K.~{Misselt},
  K.~{Okumura}, T.~{Beck}, J.~P. {Bernard}, A.~{Bolatto}, C.~{Bot}, M.~{Boyer},
  S.~{Bracker}, L.~R. {Carlson}, G.~C. {Clayton}, C.~H.~R. {Chen},
  E.~{Churchwell}, Y.~{Fukui}, M.~{Galametz}, J.~L. {Hora}, A.~{Hughes},
  R.~{Indebetouw}, F.~P. {Israel}, A.~{Kawamura}, F.~{Kemper}, S.~{Kim},
  E.~{Kwon}, B.~{Lawton}, A.~{Li}, K.~S. {Long}, M.~{Marengo}, S.~C. {Madden},
  M.~{Matsuura}, J.~M. {Oliveira}, T.~{Onishi}, M.~{Otsuka}, D.~{Paradis},
  A.~{Poglitsch}, D.~{Riebel}, W.~T. {Reach}, M.~{Rubio}, B.~{Sargent},
  M.~{Sewi{\l}o}, J.~D. {Simon}, R.~{Skibba}, L.~J. {Smith}, S.~{Srinivasan},
  A.~G.~G.~M. {Tielens}, J.~T. {van Loon}, B.~{Whitney}, P.~M. {Woods},
  {HERschel Inventory of The Agents of Galaxy Evolution (HERITAGE): The Large
  Magellanic Cloud dust}, Astronomy and Astrophysics 518 (2010) L71.

\bibitem{Bot2010}
C.~{Bot}, N.~{Ysard}, D.~{Paradis}, J.~P. {Bernard}, G.~{Lagache}, F.~P.
  {Israel}, W.~F. {Wall}, {Submillimeter to centimeter excess emission from the
  Magellanic Clouds. II. On the nature of the excess}, Astronomy and Astrophysics 523 (2010) A20.

\bibitem{Mason2020}
B.~{Mason}, S.~{Dicker}, S.~{Sadavoy}, S.~{Stanchfield}, T.~{Mroczkowski},
  C.~{Romero}, R.~{Friesen}, C.~{Sarazin}, J.~{Sievers}, T.~{Stanke},
  M.~{Devlin}, {Confirmation of Enhanced Long-wavelength Dust Emission in OMC
  2/3},  Astrophysical Journal 893~(1) (2020) 13.

\bibitem{Grun2019}
E.~{Gr{\"u}n}, H.~{Kr{\"u}ger}, R.~{Srama}, {The Dawn of Dust Astronomy}, Space Science Reviews
  215~(7) (2019) 46.

\bibitem{Koschny2019}
D.~{Koschny}, R.~H. {Soja}, C.~{Engrand}, G.~J. {Flynn}, J.~{Lasue}, A.-C.
  {Levasseur-Regourd}, D.~{Malaspina}, T.~{Nakamura}, A.~R. {Poppe}, V.~J.
  {Sterken}, J.~M. {Trigo-Rodr{\'\i}guez}, {Interplanetary Dust, Meteoroids,
  Meteors and Meteorites}, Space Science Reviews 215~(4) (2019) 34.

\bibitem{Dones2004}
L.~{Dones}, P.~R. {Weissman}, H.~F. {Levison}, M.~J. {Duncan}, {Oort Cloud
  formation and dynamics}, 2004, p. 153.

\bibitem{Fitzsimmons2018}
A.~{Fitzsimmons}, C.~{Snodgrass}, B.~{Rozitis}, B.~{Yang}, M.~{Hyland},
  T.~{Seccull}, M.~T. {Bannister}, W.~C. {Fraser}, R.~{Jedicke}, P.~{Lacerda},
  {Spectroscopy and thermal modelling of the first interstellar object 1I/2017
  U1 `Oumuamua}, Nature Astronomy 2 (2018) 133--137.

\bibitem{Frisch1999}
P.~C. {Frisch}, J.~M. {Dorschner}, J.~{Geiss}, J.~M. {Greenberg},
  E.~{Gr{\"u}n}, M.~{Landgraf}, P.~{Hoppe}, A.~P. {Jones}, W.~{Kr{\"a}tschmer},
  T.~J. {Linde}, G.~E. {Morfill}, W.~{Reach}, J.~D. {Slavin}, J.~{Svestka},
  A.~N. {Witt}, G.~P. {Zank}, {Dust in the Local Interstellar Wind},  Astrophysical Journal
  525~(1) (1999) 492--516.

\bibitem{Kruger_Grun2009}
H.~{Kr{\"u}ger}, E.~{Gr{\"u}n}, {Interstellar Dust Inside and Outside the
  Heliosphere}, Space Science Reviews 143~(1-4) (2009) 347--356.

\bibitem{Kruger2015}
H.~{Kr{\"u}ger}, P.~{Strub}, E.~{Gr{\"u}n}, V.~J. {Sterken}, {Sixteen Years of
  Ulysses Interstellar Dust Measurements in the Solar System. I. Mass
  Distribution and Gas-to-dust Mass Ratio},  Astrophysical Journal 812~(2) (2015) 139.

\bibitem{Altobelli2005}
N.~{Altobelli}, S.~{Kempf}, H.~{Kr{\"u}ger}, M.~{Landgraf}, M.~{Roy},
  E.~{Gr{\"u}n}, {Interstellar dust flux measurements by the Galileo dust
  instrument between the orbits of Venus and Mars}, Journal of Geophysical
  Research (Space Physics) 110~(A7) (2005) A07102.

\bibitem{Altobelli2007}
N.~{Altobelli}, V.~{Dikarev}, S.~{Kempf}, R.~{Srama}, S.~{Helfert},
  G.~{Moragas-Klostermeyer}, M.~{Roy}, E.~{Gr{\"u}n}, {Cassini/Cosmic Dust
  Analyzer in situ dust measurements between Jupiter and Saturn}, Journal of
  Geophysical Research (Space Physics) 112~(A7) (2007) A07105.

\bibitem{Merouane2016}
S.~{Merouane}, B.~{Zaprudin}, O.~{Stenzel}, Y.~{Langevin}, N.~{Altobelli},
  V.~{Della Corte}, H.~{Fischer}, M.~{Fulle}, K.~{Hornung}, J.~{Sil{\'e}n},
  N.~{Ligier}, A.~{Rotundi}, J.~{Ryno}, R.~{Schulz}, M.~{Hilchenbach},
  J.~{Kissel}, {Cosima Team}, {Dust particle flux and size distribution in the
  coma of 67P/Churyumov-Gerasimenko measured in situ by the COSIMA instrument
  on board Rosetta}, Astronomy and Astrophysics 596 (2016) A87.

\bibitem{Love_Brownlee1993}
S.~G. {Love}, D.~E. {Brownlee}, {A Direct Measurement of the Terrestrial Mass
  Accretion Rate of Cosmic Dust}, Science 262~(5133) (1993) 550--553.

\bibitem{Galligan_Baggaley2004}
D.~P. {Galligan}, W.~J. {Baggaley}, {The orbital distribution of radar-detected
  meteoroids of the Solar system dust cloud}, Monthly Notices of the Royal Astronomical Society 353~(2) (2004) 422--446.

\bibitem{Pokorny_Brown2016}
P.~{Pokorn{\'y}}, P.~G. {Brown}, {A reproducible method to determine the
  meteoroid mass index}, Astronomy and Astrophysics 592 (2016) A150.

\bibitem{Jewitt2014}
D.~{Jewitt}, M.~{Ishiguro}, H.~{Weaver}, J.~{Agarwal}, M.~{Mutchler},
  S.~{Larson}, {Hubble Space Telescope Investigation of Main-belt Comet
  133P/Elst-Pizarro},  Astronomical Journal 147~(5) (2014) 117.

\bibitem{Brown2002}
P.~{Brown}, R.~E. {Spalding}, D.~O. {ReVelle}, E.~{Tagliaferri}, S.~P.
  {Worden}, {The flux of small near-Earth objects colliding with the Earth},
  Nature 420~(6913) (2002) 294--296.

\bibitem{ReVelle2001}
D.~O. {Revelle}, {Global infrasonic monitoring of large bolides}, in:
  B.~{Warmbein} (Ed.), Meteoroids 2001 Conference, Vol. 495 of ESA Special
  Publication, 2001, pp. 483--489.

\bibitem{Werner2002}
S.~C. {Werner}, A.~W. {Harris}, G.~{Neukum}, B.~A. {Ivanov}, {NOTE: The
  Near-Earth Asteroid Size-Frequency Distribution: A Snapshot of the Lunar
  Impactor Size-Frequency Distribution}, Icarus 156~(1) (2002) 287--290.

\bibitem{Rabinowitz2000}
D.~{Rabinowitz}, E.~{Helin}, K.~{Lawrence}, S.~{Pravdo}, {A reduced estimate of
  the number of kilometre-sized near-Earth asteroids}, Nature 403~(6766) (2000)
  165--166.

\bibitem{Bland_Artemieva2006}
P.~A. {Bland}, N.~A. {Artemieva}, {The rate of small impacts on Earth},
  Meteoritics and Planetary Science 41~(4) (2006) 607--631.

\bibitem{Halliday1989}
I.~{Halliday}, A.~T. {Blackwell}, A.~A. {Griffin}, {The Flux of Meteorites on
  the Earth's Surface}, Meteoritics 24~(3) (1989) 173.

\bibitem{Ananyeva2020}
V.~I. {Ananyeva}, A.~E. {Ivanova}, A.~A. {Venkstern}, I.~A. {Shashkova}, A.~V.
  {Yudaev}, A.~V. {Tavrov}, O.~I. {Korablev}, J.-L. {Bertaux}, {Mass
  distribution of exoplanets considering some observation selection effects in
  the transit detection technique}, Icarus 346 (2020) 113773.

\bibitem{Ceplecha1998}
Z.~{Ceplecha}, J.~{Borovi{\v{c}}ka}, W.~G. {Elford}, D.~O. {Revelle}, R.~L.
  {Hawkes}, V.~{Porub{\v{c}}an}, M.~{{\v{S}}imek}, {Meteor Phenomena and
  Bodies}, Space Science Reviews 84 (1998) 327--471.

\bibitem{Meech2017}
K.~J. {Meech}, R.~{Weryk}, M.~{Micheli}, J.~T. {Kleyna}, O.~R. {Hainaut},
  R.~{Jedicke}, R.~J. {Wainscoat}, K.~C. {Chambers}, J.~V. {Keane},
  A.~{Petric}, L.~{Denneau}, E.~{Magnier}, T.~{Berger}, M.~E. {Huber},
  H.~{Flewelling}, C.~{Waters}, E.~{Schunova-Lilly}, S.~{Chastel}, {A brief
  visit from a red and extremely elongated interstellar asteroid}, Nature
  552~(7685) (2017) 378--381.

\bibitem{Bolin2018}
B.~T. {Bolin}, H.~A. {Weaver}, Y.~R. {Fernandez}, C.~M. {Lisse},
  D.~{Huppenkothen}, R.~L. {Jones}, M.~{Juri{\'c}}, J.~{Moeyens}, C.~A.
  {Schambeau}, C.~T. {Slater}, {\v{Z}}.~{Ivezi{\'c}}, A.~J. {Connolly}, {APO
  Time-resolved Color Photometry of Highly Elongated Interstellar Object
  1I/{\textquoteleft}Oumuamua},  Astrophysical Journal 852~(1) (2018) L2.

\bibitem{Trilling2018}
D.~E. {Trilling}, M.~{Mommert}, J.~L. {Hora}, D.~{Farnocchia}, P.~{Chodas},
  J.~{Giorgini}, H.~A. {Smith}, S.~{Carey}, C.~M. {Lisse}, M.~{Werner},
  A.~{McNeill}, S.~R. {Chesley}, J.~P. {Emery}, G.~{Fazio}, Y.~R. {Fernandez},
  A.~{Harris}, M.~{Marengo}, M.~{Mueller}, A.~{Roegge}, N.~{Smith}, H.~A.
  {Weaver}, K.~{Meech}, M.~{Micheli}, {Spitzer Observations of Interstellar
  Object 1I/{\textquoteleft}Oumuamua},  Astronomical Journal 156~(6) (2018) 261.

\bibitem{Schlichting2013}
H.~E. {Schlichting}, C.~I. {Fuentes}, D.~E. {Trilling}, {Initial Planetesimal
  Sizes and the Size Distribution of Small Kuiper Belt Objects},  Astronomical Journal 146~(2)
  (2013) 36.

\bibitem{Fuentes2010}
C.~I. {Fuentes}, M.~J. {Holman}, D.~E. {Trilling}, P.~{Protopapas},
  {Trans-Neptunian Objects with Hubble Space Telescope ACS/WFC},  Astrophysical Journal 722~(2)
  (2010) 1290--1302.

\bibitem{Freese2009}
K.~{Freese}, {Review of Observational Evidence for Dark Matter in the Universe
  and in upcoming searches for Dark Stars}, in: E.~{P{\'e}contal},
  T.~{Buchert}, P.~{di Stefano}, Y.~{Copin} (Eds.), EAS Publications Series,
  Vol.~36 of EAS Publications Series, 2009, pp. 113--126.

\bibitem{Courteau2014}
S.~{Courteau}, M.~{Cappellari}, R.~S. {de Jong}, A.~A. {Dutton}, E.~{Emsellem},
  H.~{Hoekstra}, L.~V.~E. {Koopmans}, G.~A. {Mamon}, C.~{Maraston}, T.~{Treu},
  L.~M. {Widrow}, {Galaxy masses}, Reviews of Modern Physics 86~(1) (2014)
  47--119.

\bibitem{Deckers_Teiser2014}
J.~{Deckers}, J.~{Teiser}, {Macroscopic Dust in Protoplanetary
  Disks{\textemdash}from Growth to Destruction},  Astrophysical Journal 796~(2) (2014) 99.

\bibitem{Dumke2004}
M.~{Dumke}, M.~{Krause}, R.~{Wielebinski}, {Cold dust in a selected sample of
  nearby galaxies. I. The interacting galaxy NGC 4631}, Astronomy and Astrophysics 414 (2004)
  475--486.

\bibitem{Bendo2006}
G.~J. {Bendo}, B.~A. {Buckalew}, D.~A. {Dale}, B.~T. {Draine}, R.~D. {Joseph},
  J.~{Kennicutt}, Robert~C., K.~{Sheth}, J.-D.~T. {Smith}, F.~{Walter},
  D.~{Calzetti}, J.~M. {Cannon}, C.~W. {Engelbracht}, K.~D. {Gordon},
  G.~{Helou}, D.~{Hollenbach}, E.~J. {Murphy}, H.~{Roussel}, {Spitzer and JCMT
  Observations of the Active Galactic Nucleus in the Sombrero Galaxy (NGC
  4594)},  Astrophysical Journal 645~(1) (2006) 134--147.

\bibitem{Gordon2014}
K.~D. {Gordon}, J.~{Roman-Duval}, C.~{Bot}, M.~{Meixner}, B.~{Babler}, J.-P.
  {Bernard}, A.~{Bolatto}, M.~L. {Boyer}, G.~C. {Clayton}, C.~{Engelbracht},
  Y.~{Fukui}, M.~{Galametz}, F.~{Galliano}, S.~{Hony}, A.~{Hughes},
  R.~{Indebetouw}, F.~P. {Israel}, K.~{Jameson}, A.~{Kawamura},
  V.~{Lebouteiller}, A.~{Li}, S.~C. {Madden}, M.~{Matsuura}, K.~{Misselt},
  E.~{Montiel}, K.~{Okumura}, T.~{Onishi}, P.~{Panuzzo}, D.~{Paradis},
  M.~{Rubio}, K.~{Sand strom}, M.~{Sauvage}, J.~{Seale}, M.~{Sewi{\l}o},
  K.~{Tchernyshyov}, R.~{Skibba}, {Dust and Gas in the Magellanic Clouds from
  the HERITAGE Herschel Key Project. I. Dust Properties and Insights into the
  Origin of the Submillimeter Excess Emission},  Astrophysical Journal 797~(2) (2014) 85.

\bibitem{Ferrarese_Ford_2005}
L.~{Ferrarese}, H.~{Ford}, {Supermassive Black Holes in Galactic Nuclei: Past,
  Present and Future Research}, Space Science Reviews 116~(3-4) (2005) 523--624.

\bibitem{Volonteri2012}
M.~{Volonteri}, {The Formation and Evolution of Massive Black Holes}, Science
  337~(6094) (2012) 544.

\bibitem{Greene2020}
J.~E. {Greene}, J.~{Strader}, L.~C. {Ho}, {Intermediate-Mass Black Holes},
  Annual Review of Astronomy and Astrophysics  58 (2020) 257--312.

\bibitem{Miller_Colbert_2004}
M.~C. {Miller}, E.~J.~M. {Colbert}, {Intermediate-Mass Black Holes},
  International Journal of Modern Physics D 13~(1) (2004) 1--64.

\bibitem{Kormendy_Ho_2013}
J.~{Kormendy}, L.~C. {Ho}, {Coevolution (Or Not) of Supermassive Black Holes
  and Host Galaxies}, Annual Review of Astronomy and Astrophysics  51~(1) (2013) 511--653.

\bibitem{Tsuna_Kawanaka_2019}
D.~{Tsuna}, N.~{Kawanaka}, {Radio emission from accreting isolated black holes
  in our galaxy}, Monthly Notices of the Royal Astronomical Society 488~(2) (2019) 2099--2107.

\bibitem{Schodel2002}
R.~{Sch{\"o}del}, T.~{Ott}, R.~{Genzel}, R.~{Hofmann}, M.~{Lehnert},
  A.~{Eckart}, N.~{Mouawad}, T.~{Alexander}, M.~J. {Reid}, R.~{Lenzen},
  M.~{Hartung}, F.~{Lacombe}, D.~{Rouan}, E.~{Gendron}, G.~{Rousset}, A.~M.
  {Lagrange}, W.~{Brandner}, N.~{Ageorges}, C.~{Lidman}, A.~F.~M. {Moorwood},
  J.~{Spyromilio}, N.~{Hubin}, K.~M. {Menten}, {A star in a 15.2-year orbit
  around the supermassive black hole at the centre of the Milky Way}, Nature
  419~(6908) (2002) 694--696.

\bibitem{Ghez2005}
A.~M. {Ghez}, S.~{Salim}, S.~D. {Hornstein}, A.~{Tanner}, J.~R. {Lu},
  M.~{Morris}, E.~E. {Becklin}, G.~{Duch{\^e}ne}, {Stellar Orbits around the
  Galactic Center Black Hole},  Astrophysical Journal 620~(2) (2005) 744--757.

\bibitem{Gebhardt2011}
K.~{Gebhardt}, J.~{Adams}, D.~{Richstone}, T.~R. {Lauer}, S.~M. {Faber},
  K.~{G{\"u}ltekin}, J.~{Murphy}, S.~{Tremaine}, {The Black Hole Mass in M87
  from Gemini/NIFS Adaptive Optics Observations},  Astrophysical Journal 729~(2) (2011) 119.

\bibitem{Banados2018}
E.~{Ba{\~n}ados}, B.~P. {Venemans}, C.~{Mazzucchelli}, E.~P. {Farina},
  F.~{Walter}, F.~{Wang}, R.~{Decarli}, D.~{Stern}, X.~{Fan}, F.~B. {Davies},
  J.~F. {Hennawi}, R.~A. {Simcoe}, M.~L. {Turner}, H.-W. {Rix}, J.~{Yang},
  D.~D. {Kelson}, G.~C. {Rudie}, J.~M. {Winters}, {An 800-million-solar-mass
  black hole in a significantly neutral Universe at a redshift of 7.5}, Nature
  553~(7689) (2018) 473--476.

\bibitem{Dolgov2018}
A.~D. {Dolgov}, {Massive and supermassive black holes in the contemporary and
  early Universe and problems in cosmology and astrophysics}, Physics Uspekhi
  61~(2) (2018) 115.

\bibitem{Larson2023}
R.~L. {Larson}, S.~L. {Finkelstein}, D.~D. {Kocevski}, T.~A. {Hutchison}, J.~R.
  {Trump}, P.~{Arrabal Haro}, V.~{Bromm}, N.~J. {Cleri}, M.~{Dickinson},
  S.~{Fujimoto}, J.~S. {Kartaltepe}, A.~M. {Koekemoer}, C.~{Papovich},
  N.~{Pirzkal}, S.~{Tacchella}, J.~A. {Zavala}, M.~{Bagley}, P.~{Behroozi},
  J.~B. {Champagne}, J.~W. {Cole}, I.~{Jung}, A.~M. {Morales}, G.~{Yang},
  H.~{Zhang}, A.~{Zitrin}, R.~O. {Amor{\'\i}n}, D.~{Burgarella}, C.~M. {Casey},
  {\'O}.~A. {Ch{\'a}vez Ortiz}, I.~G. {Cox}, K.~{Chworowsky}, A.~{Fontana},
  E.~{Gawiser}, A.~{Grazian}, N.~A. {Grogin}, S.~{Harish}, N.~P. {Hathi},
  M.~{Hirschmann}, B.~W. {Holwerda}, S.~{Juneau}, G.~C.~K. {Leung}, R.~A.
  {Lucas}, E.~J. {McGrath}, P.~G. {P{\'e}rez-Gonz{\'a}lez}, J.~R. {Rigby},
  L.-M. {Seill{\'e}}, R.~C. {Simons}, A.~{de La Vega}, B.~J. {Weiner}, S.~M.
  {Wilkins}, L.~Y.~A. {Yung}, {Ceers Team}, {A CEERS Discovery of an Accreting
  Supermassive Black Hole 570 Myr after the Big Bang: Identifying a Progenitor
  of Massive z > 6 Quasars},  Astrophysical Journal 953~(2) (2023) L29.

\bibitem{Goulding2023}
A.~D. {Goulding}, J.~E. {Greene}, D.~J. {Setton}, I.~{Labbe}, R.~{Bezanson},
  T.~B. {Miller}, H.~{Atek}, {\'A}.~{Bogd{\'a}n}, G.~{Brammer},
  I.~{Chemerynska}, S.~E. {Cutler}, P.~{Dayal}, Y.~{Fudamoto}, S.~{Fujimoto},
  L.~J. {Furtak}, V.~{Kokorev}, G.~{Khullar}, J.~{Leja}, D.~{Marchesini},
  P.~{Natarajan}, E.~{Nelson}, P.~A. {Oesch}, R.~{Pan}, C.~{Papovich}, S.~H.
  {Price}, P.~{van Dokkum}, B.~{Wang}, J.~R. {Weaver}, K.~E. {Whitaker},
  A.~{Zitrin}, {UNCOVER: The Growth of the First Massive Black Holes from
  JWST/NIRSpec-Spectroscopic Redshift Confirmation of an X-Ray Luminous AGN at
  z = 10.1},  Astrophysical Journal 955~(1) (2023) L24.

\end{thebibliography}
\end{document}